\documentclass[aps,prd,superscriptaddress,preprintnumbers,nofootinbib]{revtex4-1}

\pdfoutput=1

\usepackage{graphics}
\usepackage[dvips]{graphicx}
\usepackage{mathrsfs}
\usepackage{amssymb}
\usepackage{amsmath}
\usepackage{verbatim}
\usepackage{float}
\usepackage{slashed}
\usepackage{bbm}
\usepackage[dvips,letterpaper,text={6.5in,9in}]{geometry}

\newcommand{\be}{\begin{equation}}
\newcommand{\ee}{\end{equation}}
\newcommand{\bea}{\begin{eqnarray}}
\newcommand{\eea}{\end{eqnarray}}

\newcommand{\gev}{\text{GeV}}
\newcommand{\tev}{\text{TeV}}
\newcommand{\met}{\slashed E_T}
\newcommand{\none}{\chi^0_1}
\newcommand{\ntwo}{\chi^0_2}
\newcommand{\xone}{\chi^{\pm}_1}
\newcommand{\pb}{\text{pb}}
\newcommand{\fb}{\text{fb}}
\newcommand{\fbinv}{\text{fb}^{-1} }
\newcommand{\mb}{\text{mb}}
\newcommand{\nb}{\text{nb}}
\newcommand{\nn}{\nonumber}
\newcommand{\gsim}{\begin{array}{c}\sim\vspace{-21pt}\\> \end{array}}
\def\lsim{\:\raisebox{-0.5ex}{$\stackrel{\textstyle<}{\sim}$}\:}

\begin{document}
\title{Hunting Quasi-Degenerate Higgsinos}
\author{Zhenyu Han}
\affiliation{Department of Physics, University of Oregon,
             Eugene, OR 97403}

\author{Graham D. Kribs}
\affiliation{Department of Physics, University of Oregon,
             Eugene, OR 97403}
\affiliation{School of Natural Sciences, Institute for Advanced Study, 
             Princeton, NJ 08540}
\author{Adam Martin}
\affiliation{Department of Physics, University of Notre Dame, Notre Dame, IN 46556,
             USA\,}

\author{Arjun Menon}
\affiliation{Department of Physics, University of Oregon,
             Eugene, OR 97403}

\begin{abstract}
We present a new strategy to uncover light, quasi-degenerate Higgsinos, a likely ingredient in a natural supersymmetric model. Our strategy focuses on Higgsinos with inter-state splittings of $O(5-50)\,\gev$ that are produced in association with a hard, initial state jet and decay via off-shell gauge bosons to two or more leptons and missing energy, $pp \to j + \slashed E_T + 2^+\, \ell$. The additional jet is used for triggering, allowing us to significantly loosen the lepton requirements and gain sensitivity to small inter-Higgsino splittings. Focusing on the two-lepton signal, we find the seemingly large backgrounds from diboson plus jet, $\bar tt$ and $Z/\gamma^* + j$ can be reduced with careful cuts, and that fake backgrounds appear minor. For Higgsino masses $m_{\chi}$ just above the current LEP II bound ($\mu \simeq 110\, \gev$) we find the significance can be as high as $3\,\sigma$ at the LHC using the existing $20\, \fbinv$ of 8 TeV data. Extrapolating to LHC at 14 TeV with $100\,\fbinv$ data, and as one example $M_1 = M_2 = 500\, \gev$, we find $5\,\sigma$ evidence for $m_{\chi} \lesssim 140\, \gev$ and $2\,\sigma$ evidence for $m_{\chi} \lesssim 200\,\gev$. We also present a reinterpretation of ATLAS/CMS monojet bounds in terms of degenerate Higgsino ($\delta m_{\chi} \ll 5\,\gev$) plus jet production. We find the current monojet bounds on $m_{\chi}$ are no better than the chargino bounds from LEP II. 
\end{abstract}

\maketitle

\section{Introduction}
\label{sec:intro}

Higgsinos, the superpartners of the Higgs doublets, are a key element in a natural supersymmetric model. The Higgsino mass is controlled by the $\mu$ parameter which, via supersymmetry, directly enters into the tree-level mass matrix for the Higgs bosons $M^2_H$. In order for electroweak symmetry breakdown (EWSB) to occur at the correct scale without unnatural cancellations, the $\mu$ parameter must lie at the weak scale, $\sim 100\, \gev$~\cite{snatural}. The Higgs mass matrix is also influenced by other supersymmetric particles -- squarks, gluinos, winos, etc., but the Higgsinos are the only superpartners whose effect on $M^2_H$ enters at tree-level. The fact that Higgsinos play such an important role in the delicate process of EWSB in supersymmetric theories makes them a desirable target at the LHC. Studying Higgsinos at a hadron collider, however, is easier said than done.

One way to produce Higgsinos is to produce more massive, strongly coupled particles (i.e. squarks) that subsequently decay to Higgsinos. The benefits of this approach are the larger cross section for colored objects and the fact that there are lots of different colored sparticles to produce which can decay to Higgsinos (resulting in a multiplicity factor, essentially). The downside of this method is that it depends on the details of the supersymmetry spectrum through the masses and branching fractions of the colored sparticles. Furthermore, the easiest colored sparticles to produce are the first generation squarks\footnote{Gluinos production can also be large, but gluinos do not talk to electroweakinos directly.}, but they couple weakly to Higgsinos due to their small Yukawa couplings.  The amount of information light-flavor squarks can yield on the Higgsinos depends on the mixing among $\mu$, $M_1$, and $M_2$. A further problem with using light-flavor squarks as a Higgsino source is that they play little role on the Higgs sector and can therefore naturally have masses well beyond the reach of the LHC. 

Unlike light-flavor squarks, top squarks do couple strongly to Higgsinos, but they are more difficult to produce. The current stop bounds are roughly $700\, \gev$~\cite{Aad:2013ija,atlasstops1,atlasstops2,Chatrchyan:2013xna,cmsmultijet,ATLhadstops} assuming a massless lightest supersymmetric particle (LSP)\footnote{In Ref.~\cite{Kribs:2013lua}, we reinterpreted existing stop/sbottom searches in a setup with Higgsino-like LSP, and found the limits are not dramatically different, $m_{\tilde{t}} \gtrsim 700\, \gev$.} while the projected exclusion limits for same stop scenario after $3000\, \fbinv$ of luminosity at a 14 TeV LHC are only $\sim 1.1\,\tev$~\cite{eurostrategy}. While these extrapolations are rough and not tailored towards capturing Higgsinos from stop decays, the relatively small increase in limits given a huge increase in luminosity is a powerful indication of how hard top squarks are to produce and detect. Clearly, at masses not much higher than the current limits, top squarks cease to be a useful source for Higgsinos and we must look for directly produced Higgsinos instead.

Direct Higgsino production has the benefit that it is much less sensitive to the details of the rest of the spectra. However, direct production of any electroweakinos (wino, bino, Higgsinos) occurs through weak interactions, so the rates are much smaller than the production of colored superpartners. Traditionally, directly produced electroweakinos are searched for in trilepton plus missing energy final states, $3\ell + \slashed E_T$ (which we will refer to simply as trilepton searches). Trilepton searches target heavier electroweakinos that decay to the lightest electroweakino by emitting a $W/Z$.\footnote{Leptons can also be produced from chargino/neutralino decays to sleptons, but this requires that the sleptons are lighter than the electroweakinos. Our focus is on spectra with electroweakinos much lighter than all other superpartners, so we will ignore the slepton possibility here.} Typically, the largest contributing process is $ pp \to \chi^{\pm}_1\chi^0_2 \to 3\ell + \slashed E_T$. A drawback to the trilepton search is that it is only sensitive when there is a large $O(m_W, m_Z)$ splitting among the electroweakinos. As the inter-state splitting decreases, the intermediate gauge bosons in the decay chain go off-shell and the leptons they emit become too soft to trigger on efficiently; the electroweakino signal is simply lost.

Both ATLAS and CMS have searched in the trilepton channel using a ``simplified model'' approach~\cite{cmstri, atlastri}. These searches have cut a swath through parameter space\footnote{ATLAS and CMS collaborations both use lepton triggers for these events. The CMS analysis uses a dilepton trigger, requiring $p_T > 17\, \gev$ for the leading lepton, and $p_T > 8\, \gev$ for the subleading lepton. The ATLAS analysis uses a combination of single- and double-lepton triggers, with thresholds depending on the lepton flavor: $p_{T, \ell} > 25\,\gev$ for single leptons, $p_{T, \ell_1} > 25\,\gev,\, p_{T, \ell_2} > 10\, \gev$ ($p_{T, \ell_i} > 14\, \gev) $ for the asymmetric (symmetric) di-electron trigger, $p_{T, \ell_1} > 18\,\gev,\, p_{T, \ell_2} > 10\, \gev$ ($p_{T, \ell_i} > 14\,\gev) $ for the asymmetric (symmetric) di-muon trigger, and $p_{T, e} > 14\, \gev, p_{T, \mu} > 10\,\gev$ ($p_{T,\mu} > 18\,\gev, p_{T,e} > 10\, \gev)$ for the mixed di-lepton trigger. At the analysis level, both ATLAS and CMS require $p_{T, \ell} > 10\, \gev$ for all three leptons in the event.}, though, as expected from the argument above the limits rapidly degrade as the inter-electroweakino splitting drops below $\sim m_W$. This limitation is set by kinematics and is not dependent on the composition of the LSP or nearby electroweakinos. The insensitivity to small splitting will not be easily remedied with more data or higher energy. Projections from ATLAS and CMS collaborations~~\cite{eurostrategy, atlinoextrapolation, CMS:2013xfa}, albeit preliminary, show the same blind spot that is present in the existing limits. 

This blind spot exactly corresponds to the electroweakino spectrum one expects in natural supersymmetry, provided the gravitino is not the LSP.  There, the Higgsino must be light from the naturalness arguments presented earlier, while the wino and bino masses ($M_2$ and $M_1$, respectively) can be much heavier. The result is a dominantly-Higgsino LSP, accompanied by one charged and one neutral state, both $O(m^2_W/M_1, m^2_W/M_2) \sim 5-50\, \gev$ heavier than the LSP. The push for natural supersymmetry -- in light of the ever-increasing bounds on the first and second generation squark (and gluino) -- combined with the problematically-split electroweakino sectors these models possess, make new search strategies for nearly degenerate Higgsinos a high priority as we prepare for the 14 TeV LHC run. Even neglecting the UV motivation for $\mu \ll M_1, M_2$, degenerate Higgsino searches are well-motivated simply because there is no LHC bound: there are bounds approaching $800\, \gev$ on first and second generation squarks in the limit of a heavy gluino~\cite{Chatrchyan:2013lya,ATLASmsqmglu} and $\sim 1.7\, \tev$ if $m_{\tilde Q} \sim M_3$~\cite{ATLASmsqmglu}, stop/sbottom squark bounds are similar $m_{\tilde t}\sim m_{\tilde b} \sim 700\, \gev$~\cite{Aad:2013ija, atlasstops1,atlasstops2,Chatrchyan:2013xna,cmsmultijet, Chatrchyan:2013lya,ATLhadstops}, and gluinos produced in a heavy squark limit must be heavier than $1.3\, \tev$~\cite{ATLASgluino,Chatrchyan:2013iqa,ATLASmsqmglu}.  Meanwhile, as we will show, the best limit on degenerate Higgsinos still comes from LEP~II, $m_{\chi} \gtrsim 103\, \gev$~\cite{lepchargino_lim}.  If the Higgsino is absolutely stable, the lightest neutral Higgsino could be dark matter.  Given the mass hierarchy we consider, $\mu \ll M_1, M_2$, the annihilation rate is large, causing the thermal abundance of the Higgsino to be well below the cosmological abundance.  While safe from cosmological bounds, there are many ways to obtain a higher abundance without significantly affecting our signal. 

In this paper we propose a new search aimed directly at quasi-degenerate Higgsinos. Unlike the trilepton search, our search targets Higgsinos that are produced in association with a high-$p_T$ jet: $pp \to \chi\chi+j$, where $\chi$ is any state in the Higgsino multiplets. By producing this final state rather than Higgsinos alone, we have another object in the event that can be triggered upon. Using the hard initial-state radiation (ISR) for triggering, we gain the freedom to significantly relax the lepton energy requirements and push to smaller splittings, first explored in Ref.~\cite{Giudice:2010wb}. Some other recent, alternative studies on electroweakinos can be found in Ref.~\cite{Gori:2013ala, Han:2013kza, Buckley:2013kua}, though these studies do not use the $\chi\chi+j$ channel in the manner we propose.

The layout of the rest of this paper is as follows: we begin with an exploration of the parameter space and properties of nearly degenerate Higgsinos in Sec.~\ref{sec:params}. Next, in Sec.~\ref{sec:limits}, we detail our strategy using a monojet final state by looking for two soft leptons, $pp \to j + \slashed E_T + \ell\ell$. The backgrounds for this process are sizable at first, though they can be reduced with cuts. We exploit the fact that the signal leptons are softer than the backgrounds leptons, which come predominantly from the decays of on-shell gauge bosons, yielding a distinct $m_{\ell\ell}$ spectrum. Given the small signal and low-$p_T$ objects we are targeting, Sec.~\ref{sec:fakes} is devoted to a study of fake backgrounds, both from jets that fake leptons and double-parton scattering events that mimic a single hard collision. In Sec.~\ref{sec:monoj}, we reinterprete the monojet plus missing energy signal into a bound on Higgsino production (the details of our reinterpretation are given in Appendix~\ref{app:search}).  We find the reinterpreted LHC bound is no better than the LEP II bound, though it would be interesting to pursue an optimized search at the 14 TeV LHC~\cite{Han:2013usa}. Finally, we conclude with a discussion in Sec.~\ref{sec:disc}. \\

\section{The parameter space of degenerate Higgsinos}
\label{sec:params}

We wish to study supersymmetric scenarios where the Higgsino multiplets, consisting of two neutral Weyl fermions and one electrically charged Dirac fermion, are much lighter than the other electroweak fermionic superpartners (the winos and the bino). In terms of supersymmetry mass parameters, this means we are interested in the hierarchy
\begin{equation}
\mu \ll M_1, M_2.
\end{equation}
As our focus is entirely on the electroweakino sector of supersymmetric theories, we will assume throughout this work that all other superpartners -- the squarks, sleptons, heavy Higgses, and gluino -- are effectively decoupled. 
Once electroweak symmetry is broken, the Higgsinos mix with the neutral bino, and both the charged and neutral components of the wino multiplet. This mixing splits the Higgsino multiplets, giving slightly different masses to the two neutral Higgsino states $\chi_1^0$, $\chi_2^0$ and the charged state $\chi^{\pm}_1$ and endowing these three states with a small wino or bino component. The size of the splitting and the hierarchy among the three states depends on the size of $M_1$ and $M_2$ relative to each other and to $\mu$. Note that, had we chosen the bino or wino to be the lightest electroweakino, the number of light states would be different;  one neutral state for a light bino, or one neutral and one charged state for a light wino. 

To get some idea for the typical splitting size and parametric dependences of the mass splitting, we first proceed analytically and look in two simple limits, $M_1 \gg M_2 > |\mu|$ and $M_2 \gg M_1 > |\mu|$. Throughout this study we will take $M_1$ and $M_2$ to be strictly positive. The neutralino mass matrix in the MSSM in the $(\widetilde{B}^0, \widetilde{W}^0, \psi_d^0, \psi_u^0)$ basis is
\bea
M_{\tilde N^0} = \left(\begin{array}{cccc}
M_1 & 0 & - m_W t_{\theta_W} c_\beta & m_W t_{\theta_W} s_\beta \\
0 & M_2 & m_W c_\beta  & - m_W s_\beta \\
- m_W t_{\theta_W} c_\beta  & m_W c_\beta & 0 & -\mu \\
m_W t_{\theta_W} s_\beta & - m_W s_\beta & -\mu & 0
\end{array}\right) \label{eq:mNfull}
\eea
while the chargino mass matrix is
\bea
M_{\tilde C} = \left(\begin{array}{cc}
M_2 & \sqrt{2} s_\beta m_W \\
\sqrt{2} c_\beta m_W & \mu 
\end{array}\right) \label{eq:mCfull}
\eea
where for simplicity we neglect all CP phases.\\

\noindent
\emph{Case I: $M_1 \gg M_2 > |\mu|$}\\
In this case, the heavy bino can be integrated out.
Depending on the sign of $\mu$, the mixing angle between $\widetilde{W}^0$ and $\psi_-^0= \frac{1}{\sqrt{2}} \left[\psi_u^0 - \psi_d^0 \right]$, or $\widetilde{W}^0$ and $\psi_+^0 = \frac{1}{\sqrt{2}} \left[\psi_u^0 + \psi_d^0 \right]$, can be enhanced as $M_2$ approaches $\pm \mu$.  When $m_W \ll  M_2 \mp \mu$, the splittings between the mostly Higgsino states are
\bea
\left|m_{\chi^\pm_1} \right| - \left|m_{\chi^0_1} \right| &\approx&  \frac{m_W^2(1\mp   s_{2\beta})}{2(M_2 
+ |\mu|)}  \label{eq:chi1pmchi10a}\\
\left|m_{\chi^0_2} \right| - \left|m_{\chi^\pm_1} \right| \approx   \frac{m_W^2(1\pm   s_{2\beta})}{2(M_2 - |\mu|)} \label{eq:chi1pmchi20a}\,,&\quad&  \left|m_{\chi^0_2}\right| - \left|m_{\chi^0_1}\right| \approx \frac{m_W^2\,(\pm|\mu|\,s_{2\beta} + M_2)}{(M^2_2 - |\mu|^2)}
\label{eq:dmeq1}
\eea
where the $\pm$ index corresponds to when $\mu$ is positive or negative, respectively. As $M_2$ approaches $|\mu|$ the wino fraction in the lightest neutralino increases, while the wino fraction in the next-to lightest neutralino remains approximately constant. From Eq.~(\ref{eq:chi1pmchi10a}) and Eq.~(\ref{eq:chi1pmchi20a}) it is clear that the lightest neutralino and chargino are more degenerate than the second neutralino and the lightest chargino. \\

\noindent
\emph{Case II: $M_1 \gg M_2 > \mu$ scenario}\\
In this case, the heavy wino component can be integrated out. 
Similar to the previous scenario, depending on the sign of $\mu$, the splittings between the lightest neutralinos and the chargino are
\bea
\left|m_{\chi^\pm_1} \right| - \left|m_{\chi^0_1} \right| &\approx&   \frac{m_W^2 t_{\theta_W}^2(1\pm s_{2\beta})}{2(M_1 - |\mu|)} \label{eq:chi1pmchi10b}\\
\left|m_{\chi^0_2} \right| - \left|m_{\chi^\pm_1} \right| \approx   \frac{m_W^2 t_{\theta_W}^2(1\mp s_{2\beta})}{2(M_1 + |\mu|)} \label{eq:chi1pmchi20b}\,, &\quad& \left|m_{\chi^0_2}\right| - \left|m_{\chi^0_1}\right| \approx \frac{m_W^2 t_{\theta_W}^2\,(\pm|\mu|\,s_{2\beta} + M_1)}{(M^2_1 - |\mu|^2)}
\label{eq:dmeq2}
\eea
Since $t_{\theta_W} \simeq 0.5$, the mixing between the (heavy) bino and the Higgsino is smaller than the mixing of a (heavy) wino and the Higgsino of the first case, leading to a smaller overall splitting between the neutralinos and the chargino. Furthermore, unlike the first case, the splitting between $\chi_1^0$ and $\chi_1^\pm$ is greater than between $\chi_2^0$ and $\chi_1^\pm$.\\

\noindent
\emph{Numerical Scan of the bino-wino Parameter Space}\\

\begin{figure*}[t!]
\includegraphics[width=0.45\textwidth]{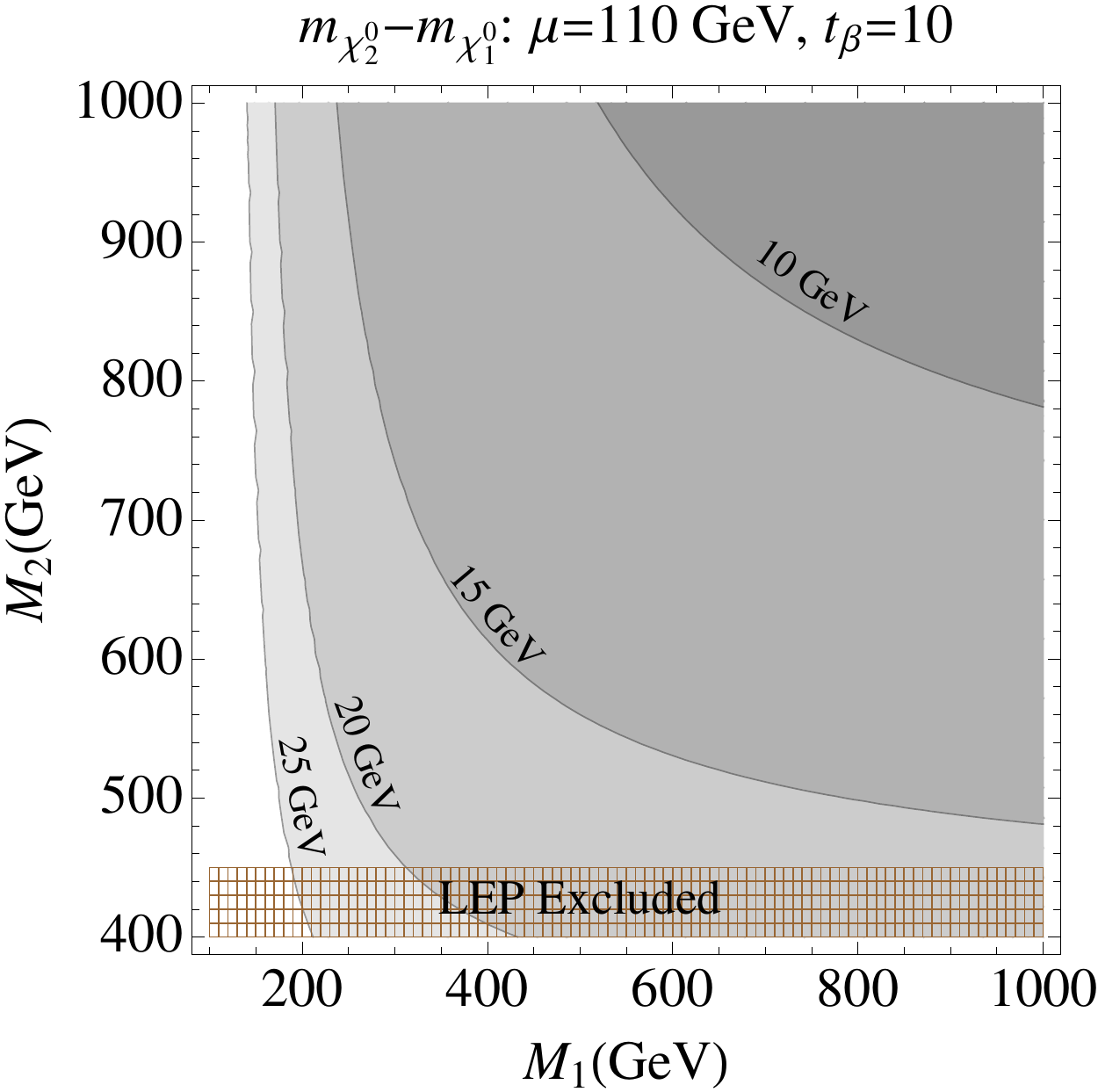}
\includegraphics[width=0.45\textwidth]{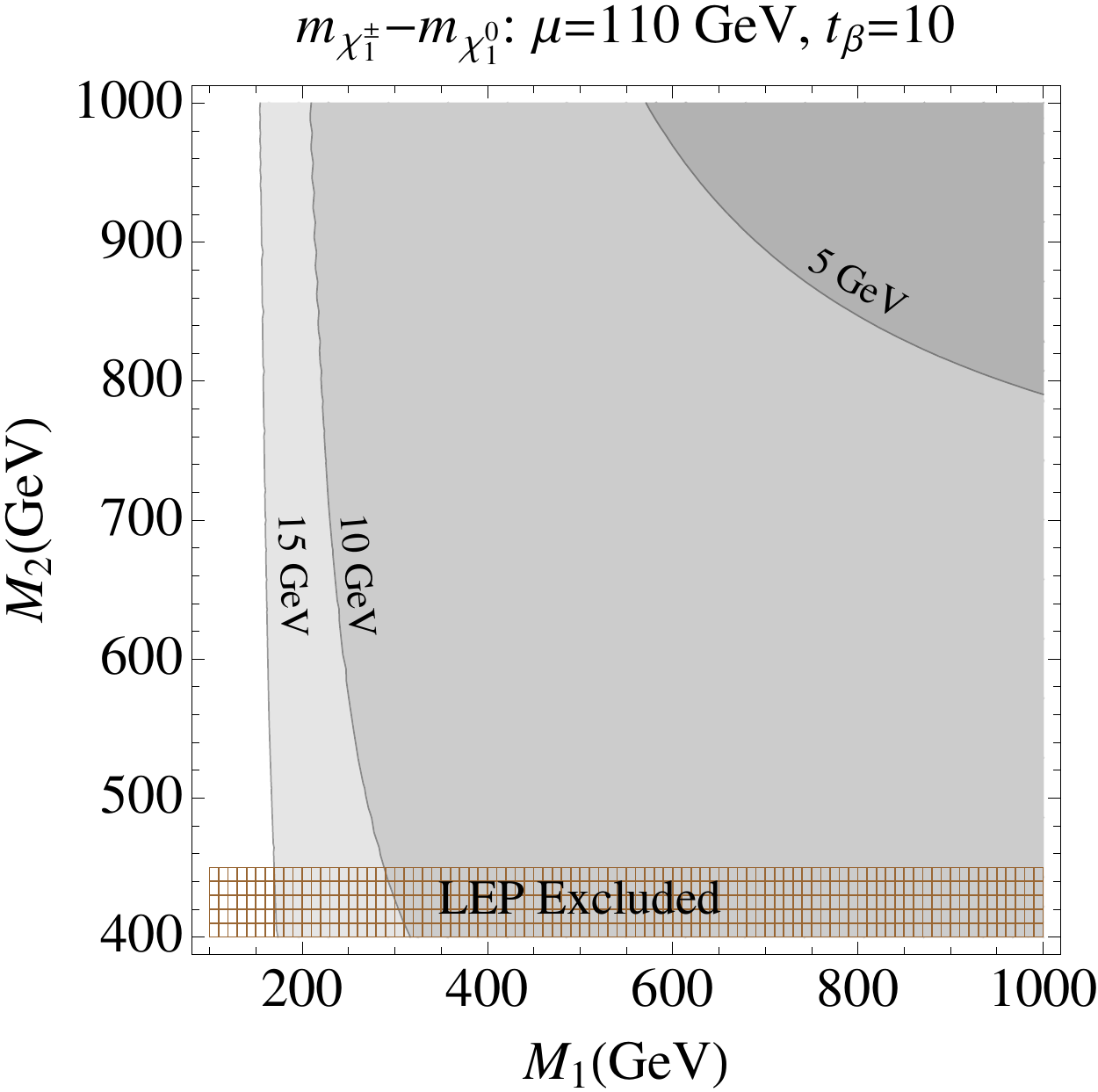} \\
\includegraphics[width=0.45\textwidth]{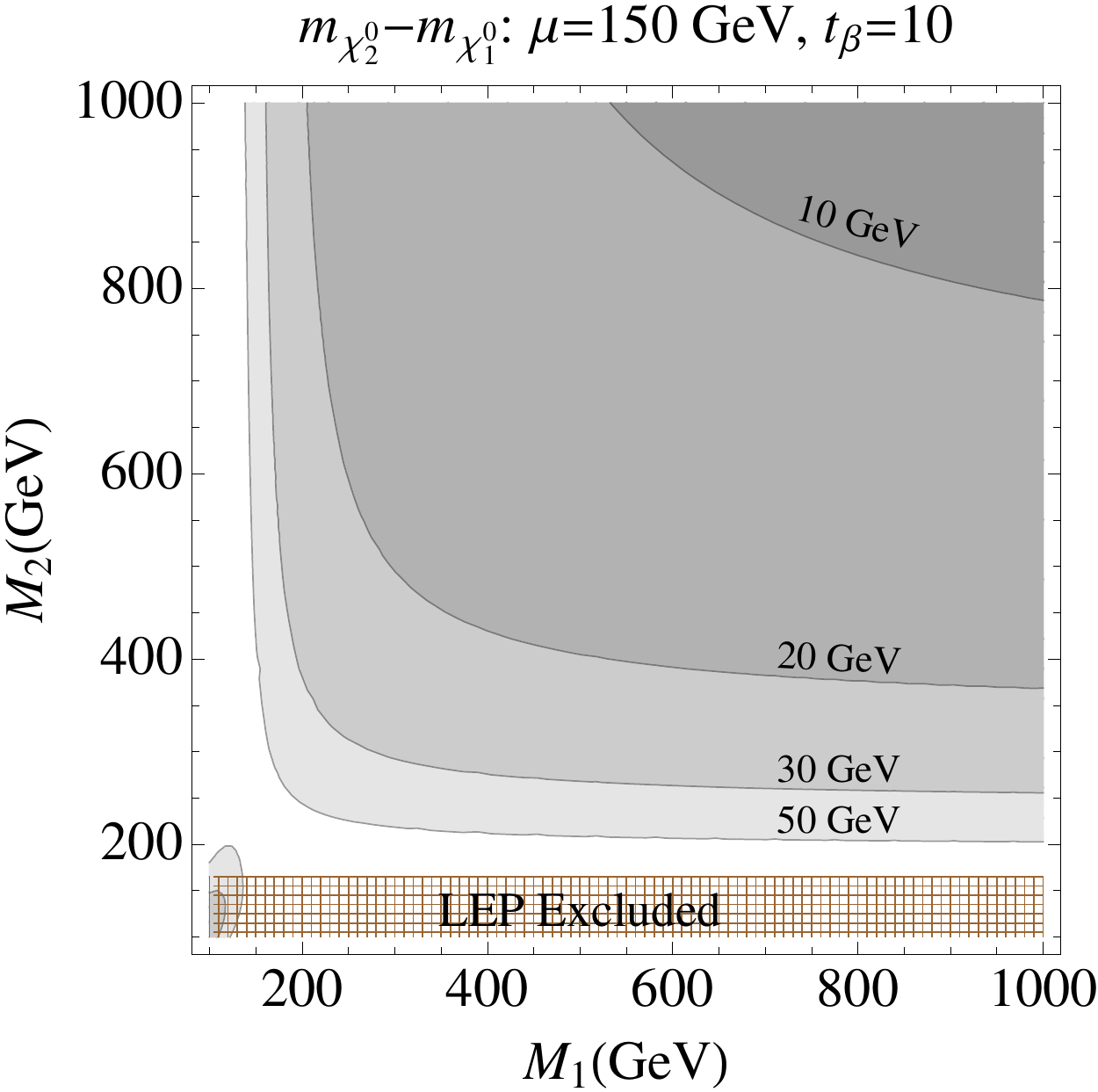}
\includegraphics[width=0.45\textwidth]{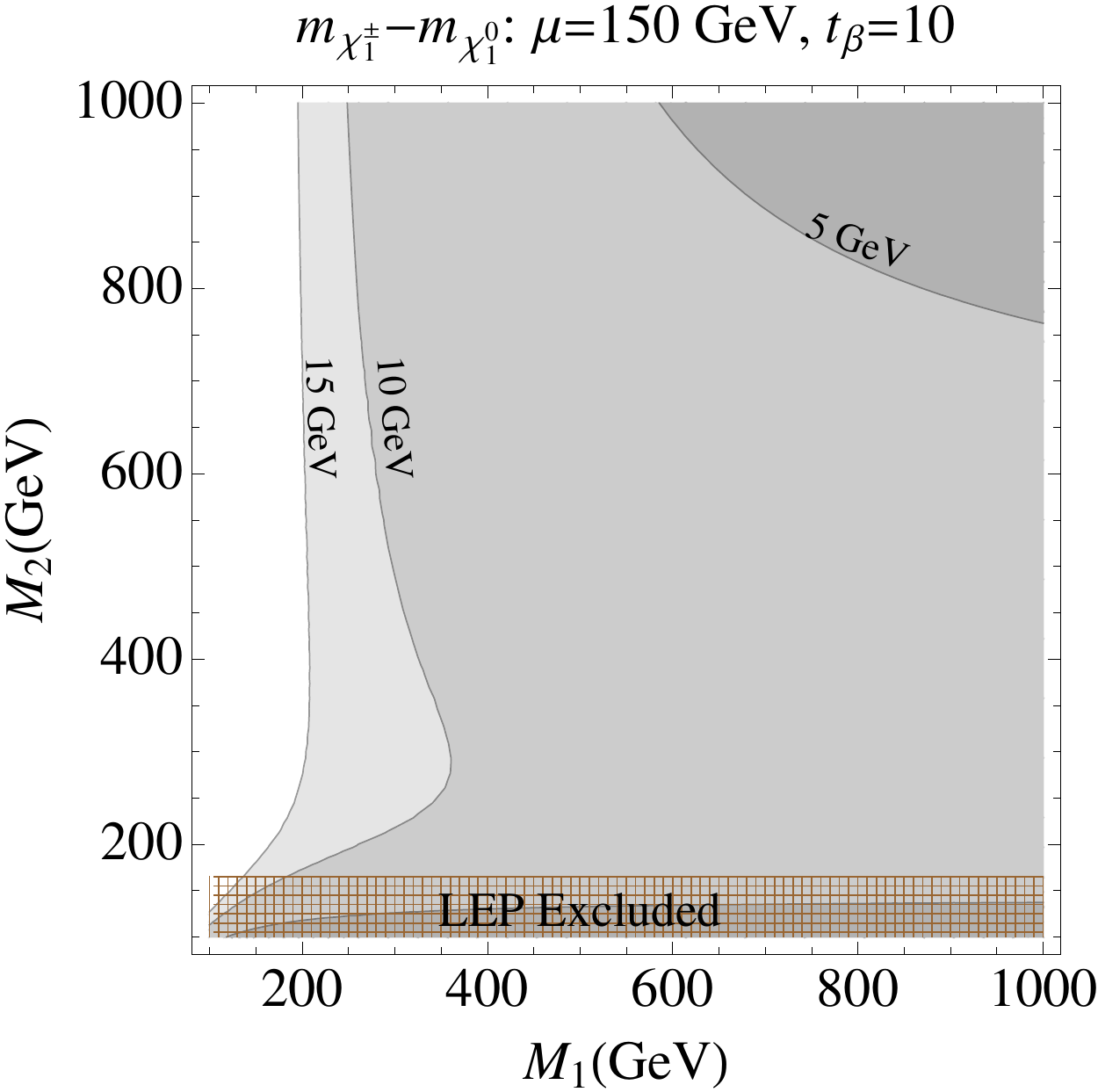}
\caption{Difference in $\ntwo - \none$ masses (left plots) 
and $\xone -\none$ masses (right plots) and 
for $\mu = 110\,\gev$ (top panels) and $\mu = 150\, \gev$ (bottom panels). In all plots, $\tan\beta = 10$.} 
\label{fig:splittingsscan}
\end{figure*}

In Fig.~\ref{fig:splittingsscan} we show the inter-Higgsino splitting more generally, as a function of both $M_1$ and $M_2$.  The regions of low $M_2$ and $\mu$ are constrained by the LEP II limit $m_{\chi_1^\pm} \gtrsim 103.5$~GeV~\cite{lepchargino_lim} when the splitting between the lightest two states is larger than $3$~GeV (that is true throughout the parameter space we consider). 

Several observations can be made from the results of Fig.~\ref{fig:splittingsscan}.  One observation is the mass hierarchy: $m_{\chi^0_2} - m_{\chi^0_1}$ is greater than $m_{\chi^{\pm}_1} - m_{\chi^0_1}$ throughout the parameter space. For $M_1, M_2 < 1\, \tev$, $m_{\chi^0_2} - m_{\chi^0_1} > 10\, \gev$, while the splitting between the lightest chargino and lightest neutralino is $> 10\,\gev$ only for $M_1, M_2 \lesssim 500\, \gev$ (for $\mu = 110\, \gev$).  As $M_1$ or $M_2$ are lowered, the splitting increases, with a steeper gradient in the $M_2$ direction. This is especially true for $m_{\chi^{\pm}} - m_{\chi^0_1}$ which, because the bino mass does not affect the chargino sector, is largely independent of $M_1$. The size and sign of $\mu$ also affects the inter-state splitting, as we can see by comparing the top and bottom panels in Fig.~\ref{fig:splittingsscan} (and from Eq.~(\ref{eq:dmeq1},\ref{eq:dmeq2})). Clearly, changes in $\mu$ have a larger effect on the inter-Higgsino splitting when either $M_1$ or $M_2$ is small.  Additionally, larger $\mu$ increases the available $M_1, M_2$ parameter space by elevating the lightest chargino above the LEP bound. Regarding the sign of $\mu$, for the $|\mu|$ and $\tan{\beta}$ values we are considering, the net effect of flipping the sign of $\mu$ is a small increase (decrease), $\lesssim 5\, \gev$ in $m_{\chi^{\pm}_1} - m_{\chi^0_1}$ ($m_{\chi^0_2} - m_{\chi^0_1}$) for all $M_1, M_2$.  Given that the shift from positive to negative $\mu$ is small, and that our results are not tied to any particular UV setup, we assume $\mu > 0$ for the remainder of this work.

Another key ingredient in Higgsino phenomenology is how the heavier $\chi^{\pm}_1$ and $\chi^0_2$ decay. For the chargino, there is only one option: $\chi^{\pm}_1 \to W^*\,\chi^0_1$. For most mass splittings we are considering, the $W^*$ branching ratios are essentially the same as for on-shell $W$. The exception is when the $\chi^{\pm} - \chi^0_1$ mass difference is below $\sim 2\,\gev$ -- where the $W^* \to \bar c s\, (c \bar s)$ and $W^* \to \tau^{\pm} \nu$ decays become kinematically squeezed, causing a slight increase in the branching fractions to lighter states. For the $\chi^0_2$ state there are three options, i.) decay to $\chi^0_1$ via an off-shell $Z^*$, ii.) decay to $\chi^{\pm}_1$ via an off-shell $W^*$, and iii.) loop level decay to a photon and $\chi^0_1$. The breakdown between the three options depends on composition of $\chi^0_2,\, \chi^0_1$, which determines the couplings, and kinematics. Since $\chi^0_2$ can decay to either sign chargino, one might expect the branching fraction to $W^*$ to be larger than $Z^*$. However, the decays to chargino are more suppressed kinematically, since we see from Fig.~\ref{fig:splittingsscan} that  $m_{\chi^{\pm}_1} - m_{\chi^0_1}$ is smaller than $m_{\chi^{0}_2} - m_{\chi^0_1}$. The $\chi^0_2$ decays to charginos are also suppressed when $M_1$ is lighter and there is a non-negligible bino component in the neutral $\chi$, since the bino does not interact with electroweak gauge bosons. This can be contrasted with the situation when $M_2$ is light. There the Higgsinos mix with the wino, an electroweak triplet that possesses stronger couplings to gauge bosons, generating a larger $\chi^0_2 \to W^* \chi^0_1$ branching fraction. These tendencies are verified in Fig.~\ref{figure:then2BR}  below, where we plot the $\chi^0_2 \to Z^*\chi^0_1$ and $\chi^0_2 \to W^*\chi^{\pm}_1$ branching ratios for $M_1, M_2 < 1\, \tev$ with $\mu = \pm 110\, \gev, \tan{\beta} = 10$. We find the two-body decay $\chi^0_2 \to \gamma \chi^0_1$ only makes up $O(1\%)$ of the total width due to the extra power of $\alpha_{em}$ it requires\footnote{The loop-factor suppression in the two-body decay is comparable to the suppression in the three-body decays from extra phase space factors, so the difference between the two- and three-body modes is primarily the extra coupling powers.}.

\begin{figure}[t]
\centering
\includegraphics[width=0.49\textwidth]{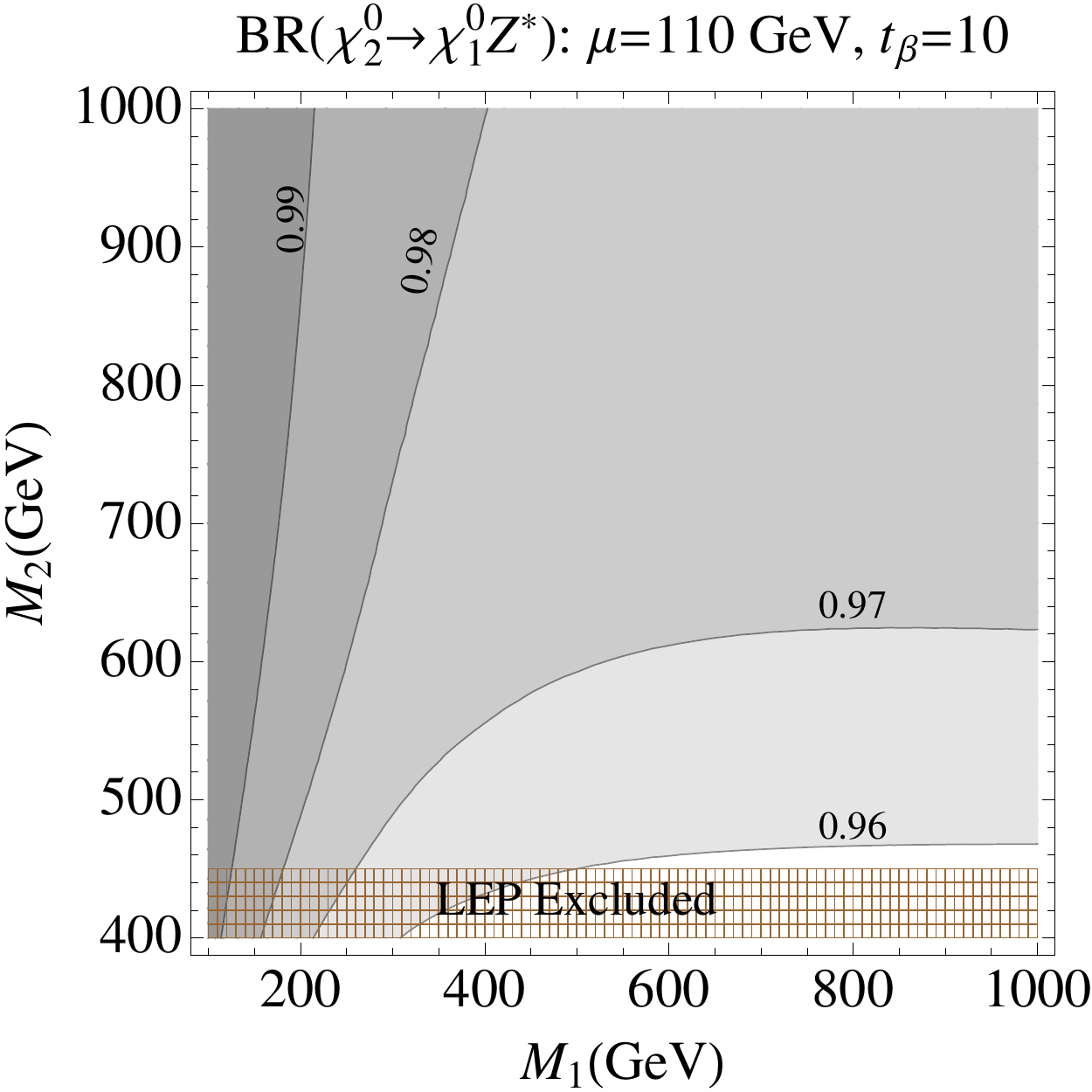}
\includegraphics[width=0.49\textwidth]{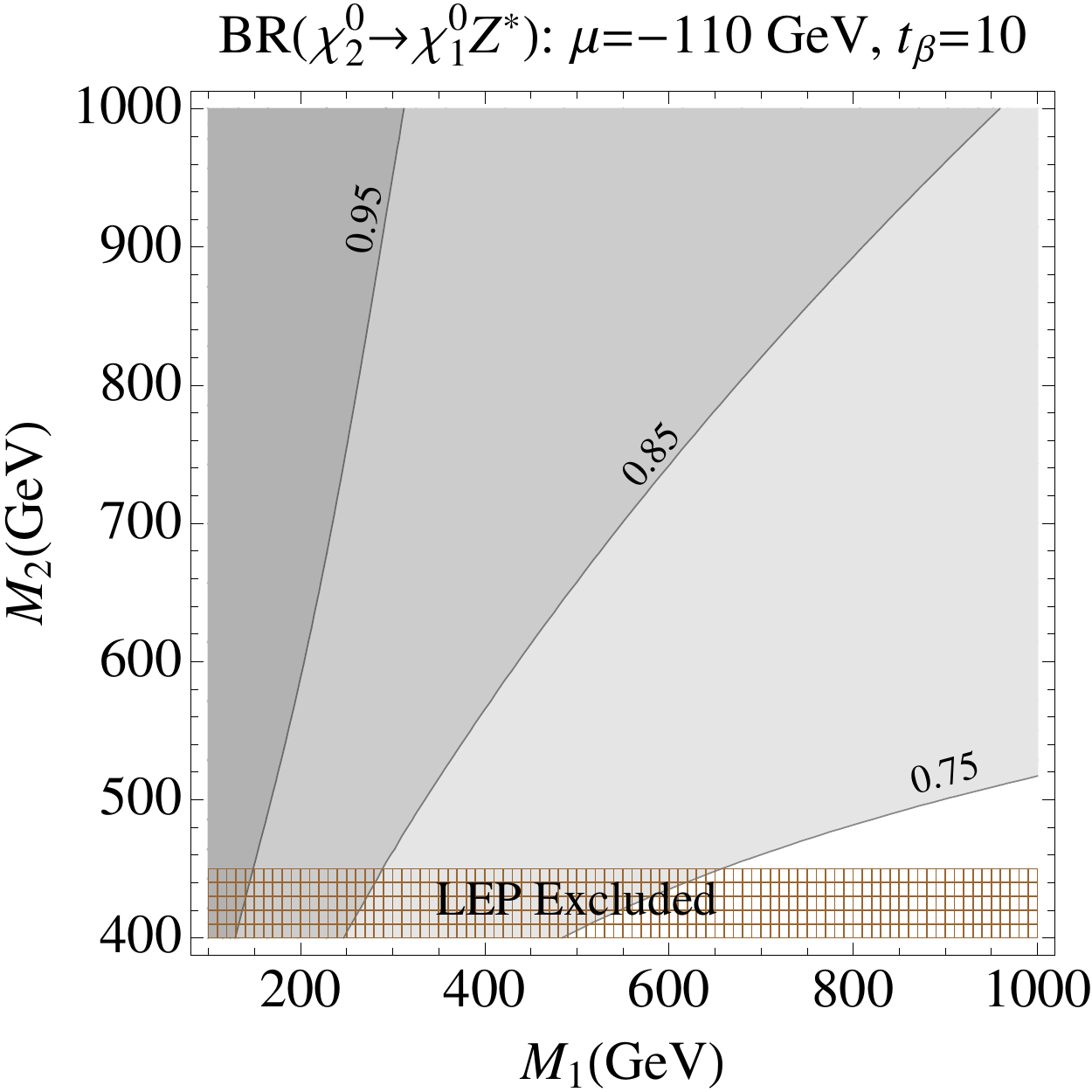}
\caption{Contours of the branching ratio $\chi^0_2 \to Z^*\chi^0_1$ for $\mu = \pm 110\,\gev$ in the $(M_1, M_2)$ parameter space. The branching fraction $\chi^0_2 \to W^*\chi^{\pm}_1$ is approximately given by $1-BR(\chi^0_2 \to Z^*\chi^0_1)$, up to a small $O(1\%)$ branching fraction for $\chi^0_2 \to \gamma \chi^0_1$.}
\label{figure:then2BR}
\end{figure}

We have seen that the inter-Higgsino mass splittings $m_{\chi_2^0} - m_{\chi_1^0}, m_{\chi^{\pm}_1} - m_{\chi_1^0}$ are $O(10\, \gev)$ for a wide range of $M_1, M_2 > |\mu|$ parameter space. Splittings of this size are substantially smaller than what the trilepton searches are sensitive to, so all three states $\chi_1^0, \chi_2^0, \chi^{\pm}_1$ are effectively invisible for this search -- an `LSP multiplet'. At the same time, the mass splittings are large enough that the $\chi^0_2$ and $\chi^{\pm}_1$ decays are still prompt. Thus, we return to the question posed earlier: with trilepton searches insensitive, how do we directly look for nearly degenerate Higgsinos? One way to proceed is to continue trilepton searches but to look for the heavier electroweakinos, the mostly-bino or mostly-wino states with mass $\sim M_1$ or $\sim M_2$, respectively. These states are heavy, so the leptons from their cascade-decays down to the LSP multiplet will carry sufficient energy to efficiently pass analysis and trigger cuts. However, there are many subtleties one must consider when extrapolating trilepton searches to the scenario we are studying here. To understand the subtleties, first consider the situation where the lightest state above the Higgsino multiplet is mostly bino. Since it is neutral, we will denote this state as $\chi_3^0$. A pure bino state does not interact with electroweak gauge bosons, so production at the LHC must proceed via the small Higgsino/wino component of $\chi_3^0$. This mixing renders $\chi^0_3$ pair-production completely negligible, leaving associated production $\chi_3^0\chi_2^0,\,  \chi_3^0\chi_1^0,\, \chi_3^0\chi_1^{\pm}$ as the only possibility. None of these associated production modes, however, lead to a trilepton signal; the $\chi_1^0,\, \chi^0_2,\, \chi^{\pm}_1$ arm of the production is invisible, while the $\chi_3^0$ can only produce one ($\chi_3^0 \to W^{\pm}(\ell\nu)\chi^{\pm}_1$) or two ($\chi_3^0 \to Z(\ell\ell)\chi^{0}_1,Z(\ell\ell)\chi^{0}_2 $) leptons. The situation is different when the state above the Higgsino multiplet is a wino. In this case, there is an extra charged state $\chi^{\pm}_2$ accompanying the neutral state $\chi_3^0$, and both states possess full-strength (not suppressed by mixing) couplings to electroweak gauge bosons. In this case, trilepton signals are possible, i.e. $pp \to \chi^0_3\chi_2^{\pm} \to W(\ell\nu)Z(\ell\ell)\chi^0_1\chi^0_2$, but the branching ratio is not straightforward. One complication is that $\chi_1^{\pm}$ is effectively invisible, so some $\chi^0_3$ decays only give one lepton rather than two, and some $\chi^{\pm}_2$ decays give two leptons rather than one. A more substantial complication is that $\chi_3^0,\, \chi^{\pm}_2$ decays to Higgs bosons, coming from the Higgsino-wino-Higgs vertices, become important. Since the majority of Higgs decays are hadronic, any $\chi_3^0,\, \chi^{\pm}_2$ decays to a Higgs boson reduces the trilepton rate. This role of Higgs decays from heavier electroweakinos, including prospects for using the Higgs decays as a discovery mode, was discussed recently in Ref.~\cite{Han:2013kza}. Given these various subtleties, it is far from clear how much of the degenerate Higgsino parameter space can be carved out by trilepton searches for the heavier states ($\chi_3^0, \chi^{\pm}_2$), even at a $14\, \tev$ LHC and high luminosity. Alternative strategies are thus warranted for all $|\mu| \ll M_1, M_2$ values. In the following sections we will forget the heavier ($\chi^0_3, \chi^{\pm}_2$) electroweakinos and explore new ways to directly probe the Higgsino LSP multiplet.

\section{Hunting Quasi-Degenerate Higgsinos:  $pp \to j + \slashed E_T + \ell \ell$}
\label{sec:limits}

In this section we exploit the small, but nonzero splitting between $\chi^{\pm}_1$, $\chi^0_2$ and $\chi^0_1$ that we outlined in the previous section. As already emphasized, the problem with small inter-Higgsino splitting is that there are no objects in the $pp \to \chi\chi$ final state that are energetic enough to trigger upon efficiently. A simple fix for this conundrum is to look at associated production $pp \to \chi\chi + X$, rather than $pp \to \chi \chi$ alone. There, the associated object $X$ can be used for triggering, turning a previously ``invisible'' event into something that can be retained for study. There are many possibilities for what triggerable object can be produced in association with Higgsino pairs -- jets, photons, $W/Z$, etc.\  -- however $\chi\chi+j$ has the highest rate. As rate is a precious commodity when looking at electroweak-strength production, we focus on this possibility, effectively searching for a Higgsino signal in a monojet-triggered event sample. 

Monojet searches have been performed by both the ATLAS and CMS  collaborations~\cite{atlasmono, cmsmono}. Their results are usually cast in the parameter space of extra-dimensional models, and, more recently, in terms of operators controlling dark matter (DM) pair production~\cite{Birkedal:2004xn, Goodman:2010yf, Bai:2010hh, Goodman:2010ku}. There are two possibilities.  If the inter-Higgsino mass splitting $m_{\chi^0_2} - m_{\chi^0_1}, m_{\chi^{\pm}_1} - m_{\chi^0_1} \equiv \delta m_{\chi} \ll 5\, \gev$, the basic monojet $\slashed E_T + j$ signal is the only option to search for such highly degenerate Higgsinos.  We defer consideration of this possibility to Sec.~\ref{sec:monoj}, where we reinterpret the existing bounds in terms of Higgsino production.

On the other hand, when the intra-Higgsino mass splitting is larger, $\delta m_{\chi} \sim 5 - 50\, \gev$, we propose looking (offline) for the soft leptonic decay products created as heavier Higgsinos decay to the lightest state via off-shell $W/Z$. Depending on the Higgsinos that are produced, a $\chi\chi + j$ event can have between $0$ and $4$ soft leptons. While we briefly discuss the $1$ and $3$ lepton possibilities later in Sec.~\ref{sec:other}, here we focus on the signal with $2$ isolated leptons $pp \to j + \slashed E_T + \ell\ell$.

There are several possible Higgsino production processes that can contribute to the $j + \slashed E_T + \ell\ell$ final state:  
\begin{align}
\label{eq:allinoproc}
1.)\,  & p p \rightarrow \chi_1^{\pm} \chi_1^{\mp}+j \to \ell^{+}\ell'^- \nu\bar{\nu}\chi^0_1\chi^0_1 + j \nn \\
2.)\, & p p \rightarrow \chi_2^{0} \chi_1^0+j \to \ell^+\ell^-  \chi^0_1 + j \nn \\
3.)\, & p p \rightarrow \chi_1^{\pm} \chi_2^{0}+j \to \ell^{+}\ell^- jj \chi^0_1\chi^0_1 + j, \,  \ell^{+}\ell^- \ell'^{\pm}\nu \chi^0_1\chi^0_1 + j, 
\end{align}
where we have omitted $\ntwo\ntwo + j$ production since it has a very small cross-section. By adding several subprocesses the total signal is enhanced, though the actual enhancement depends on the cuts imposed and on the size of the $\chi^{\pm}_1 - \chi^0_1$ and $\chi^0_2 - \chi^0_1$ mass splittings.  The dominant backgrounds for this final state are: $t\bar t$, $Z/\gamma^*(\tau^+\tau^-) + j$ with both taus decaying leptonically, and diboson plus jet.  All diboson plus jet processes that contribute to $\ell\ell+\slashed E_T +j$ are included ($W^+(\ell\nu)W^-(\ell\nu) + j, Z(\bar{\nu}\nu)Z(\ell\ell)+ j$, and $Z(\bar{\nu}\nu)\gamma(\ell\ell)+ j$), though $WW+j$ is by far the dominant contribution.

To study and compare the signal and backgrounds, we turn to Monte Carlo. We simulate the hard processes for the signal in Eq.~(\ref{eq:allinoproc}) and the major backgrounds with Madgraph 5~\cite{Alwall:2011uj}\footnote{We generate events using the following parton-level cuts: $p_{T,j} > 80\, \gev, |\eta_j| < 5.0, \slashed E_T > 80\, \gev$, $p_{T,\ell} > 5\, \gev, |\eta_{\ell}| < 2.5, $ and $\Delta R_{\ell-\ell} > 0.1$. We set $\Delta R_{j-\ell} > 0.4$ for all backgrounds except $W + \gamma^* + j$, where we use $\Delta R_{j-\ell} > 0.1$. These cuts are slightly softer than the analysis-level cuts.}. The neutralinos and charginos are fully decayed in Madgraph 5, and therefore the spin correlation is always retained. The background processes include  $j\ell^+\ell^-\nu\bar{\nu}$ (dominated by $WW$+jet), $j\tau\tau$ in the dileptonic decay channels (dominated by $j+Z, Z\rightarrow \tau\tau$) and $t\bar t$ in the dileptonic channels. The parton-level events are then showered and hadronized with Pythia 8~\cite{Sjostrand:2007gs}. In order to estimate the effect of experimental resolutions, which is important especially for the jet momentum and missing momentum measurements, we group the particles in 0.1$\times$ 0.1 calorimeter cells on the $(\eta, \phi)$ plane, roughly corresponding to the HCAL granularities. 

The event is reconstructed by first finding the isolated leptons that satisfy the following criteria: the sum of all tracks' $p_T$ within $\Delta R=0.2$ around the lepton is less than $1.8\,\gev$. No smearing is applied to the isolated leptons. The isolated leptons are then removed from the calorimeter cells which are used for jet clustering. Jet clustering is done with FastJet 3~\cite{Cacciari:2005hq, Cacciari:2011ma} using the anti-$k_t$ algorithm ($R=0.4$). After obtaining the jets, we calculate the missing transverse momentum by summing over the momenta of all isolated leptons and jets.  We then apply the following cuts, which we find effectively differentiate signal from background.

\begin{itemize}
\item large $\met$. All signal events end in two massive LSP neutralinos. When both neutralinos are forced to recoil against another hard object in the event -- a jet in the case here -- they lead to a  large $\slashed E_T$ signature. We require $\slashed E_T > 100\, \gev$.
\item exactly 1 hard jet: The topology we are interested in is a nearly invisible Higgsino system recoiling against initial-state radiation, so a single hard jet will suffice. The $t\bar t$ background, on the other hand, is characterized by at least two hard jets. By restricting the number of final state jets to a single light-flavor (anti b-tagged) jet, we can remove the vast majority of the $t \bar t $ background without affecting the signal. The diboson and $Z/\gamma^*$ backgrounds are also insensitive to the jet restriction. In practice, we require exactly one jet (jets with $p_T$ less than 30 $\gev$ are not counted), $p_{T,j} > 100\, \gev$, $|\eta_j| < 2.5$. We use $100\, \gev$ to satisfy ATLAS/CMS single-jet or, in combination with the $\slashed E_T$ cut above, the jet+$\slashed E_T$ trigger requirements, at least at the $8\, \tev$ LHC. If the jet is $b$-tagged, the event is vetoed. We will assume a $b$-tag efficiency of 80\% and neglect the possibility of light jets faking $b$'s. Reinstating $b$-fakes would have a very minor impact on our results, since the signal and the dominant $WW+ j$ background would both decrease by the same small amount. Actually, given that the jet composition of the signal and dominant background are so similar, a more aggressive tagging/fake point may be even better -- for example, a 10\% fake rate may be tolerable if we could remove 90\% of the $\bar tt$ -- but we did not attempt any such optimization in this work. 
\item two isolated leptons:   If more than two isolated leptons are found, the two leading ones are used in the following steps. We use a lower transverse momentum threshold of $7\, \gev$ (regardless of the lepton flavor), and require all leptons to lie within the tracker $|\eta_{\ell}| < 2.5$. The lower limit of  $7\, \gev$ is comparable to the off-line lepton identification thresholds in ATLAS/CMS~\cite{Chatrchyan:2012ufa,  ATLAS:2012ac, ATLAS4l}.  More precise values of the thresholds depend sensitively on the particular detector and detection region (e.g.\ in $\eta$), as well as the desired purity of the lepton sample.  This level of precision is beyond the scope of this paper, however we emphasize that the offline lepton identification thresholds are the primary limitation to considering even smaller Higgsino mass splittings (larger $M_1, M_2$). 
\item reconstructed $m_{\tau\tau} > 150\,\gev$. The dilepton plus $\met$ that arises from the $Z/\gamma^*(\tau^+\tau^-)$ background, unlike in the signal, originates from a single particle. If we assume the intermediate tau leptons from $Z/\gamma^*$ are highly energetic, we can approximate their leptonic decays as collinear. This assumption allows us to reconstruct the two pairs of tentative neutrinos in any $\tau\tau \rightarrow \ell\ell + \met$ event, and from there we can reconstruct the would-be $m_{\tau\tau}$ distribution\footnote{The collinear approximation notoriously does not work when the taus are back-to-back. As we are always interested in $Z/\gamma^* + j$, so the tau-tau system is always somewhat boosted,  this limitation is not an issue here.}. Specifically, we assume the missing energy is due to four neutrinos, each pair with momentum collinear to a lepton: $\vec p_{\nu,1} = \xi_1 \vec p_{\ell,1},\, \vec p_{\nu,2} = \xi_2 \vec p_{\ell,2}$. Using the measured $\met$ one can solve for $\xi_1,\xi_2$, which can then be used to reconstruct $p_{\nu,i}$ or $p_{\tau,i} = p_{\ell,i} + p_{\nu,i}$, which we combine to form $m_{\tau\tau}$\footnote{While the spatial momenta is scaled with $\xi_i$, we scale the energy by $|\xi_i|$ to prevent unphysical negative energy for the neutrinos if $\xi_i < 0$ (and negative energy for the parent $\tau_i$ if $\xi_i < -1$).  Large negative $\xi_i$ occur if the missing energy vector points opposite to a lepton and $|\slashed E_T| > |p_{T, \ell}|$ -- a fairly common configuration for the signal or $WW+j$ background.}.   The $Z/\gamma^*$ background should have a narrow peak in $m_{\tau\tau}$, while the signal distribution should be fairly featureless. Cutting out the $Z$ region using this variable, we can reduce the $Z/\gamma^*$ substantially -- an absolute necessity given the enormous cross section of the $jZ\rightarrow j\tau\tau$ process. In practice we find that a broad, one-sided cut $m_{\tau\tau} > 150\, \gev$ is better than a cut focused right around the $Z$-peak. The $m_{\tau\tau}$ distributions for the backgrounds and a sample signal point are illustrated in Fig.~\ref{fig:mtautau}, where we can clearly see the reconstructed $Z$ peak in the $j\tau\tau$ background. Note that we have to use a logarithmic scale for the number of events in order to see the small contributions from the signal and the other two backgrounds. 
\begin{figure}[t!]
\centering
\includegraphics[width=0.65\textwidth]{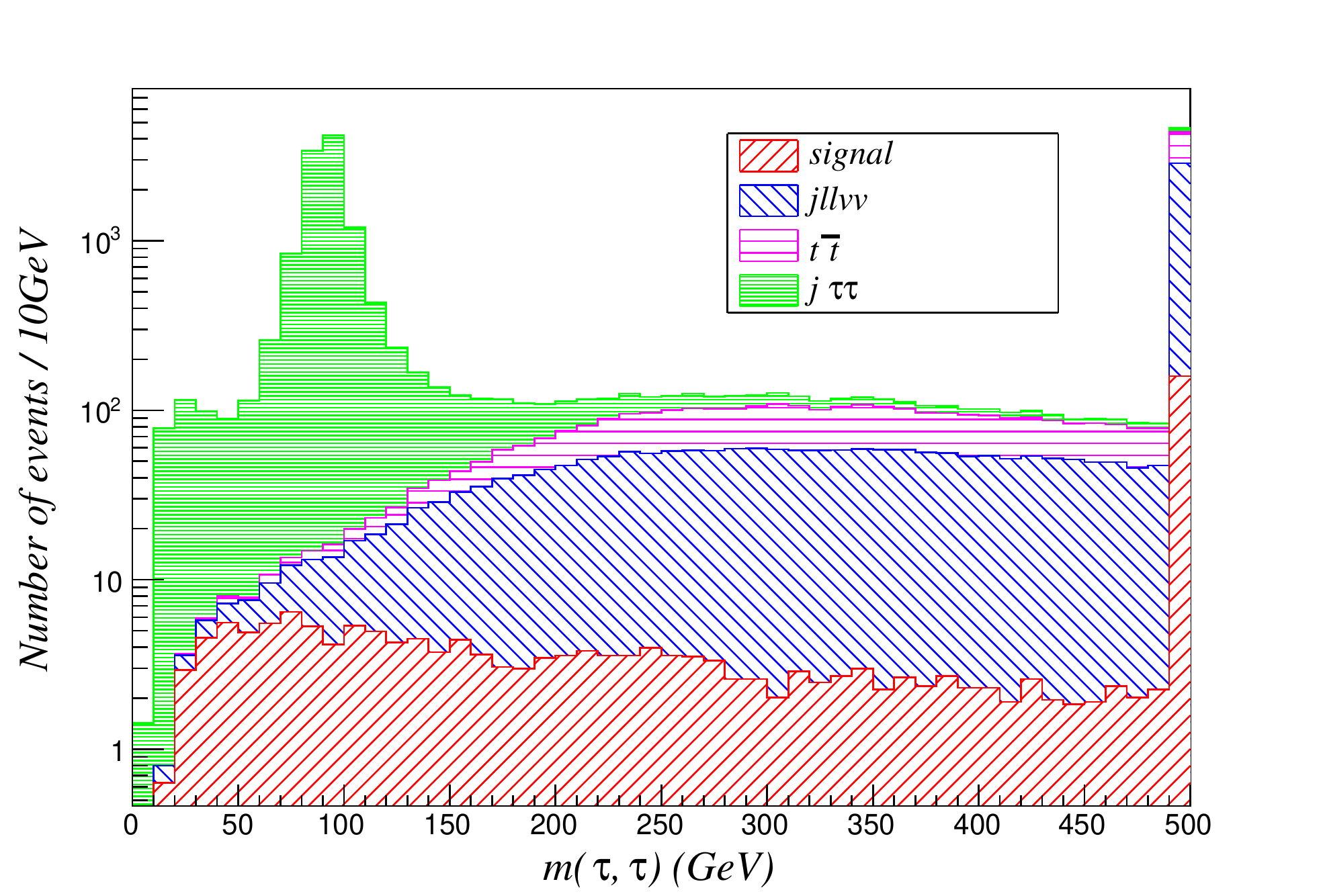}
\label{fig:mtautau}
\caption{The reconstructed $m_{\tau\tau}$, as defined in the text, for the backgrounds and a typical signal mass point: $\mu = 110\,\gev, M_1=200\,\gev, M_2=1000\,\gev$ (stacked). The last bin contains the overflow for all events with $m_{\tau\tau} > 500$~GeV.}
\end{figure}
\item Finally, we cut on the dilepton invariant mass, $m_{\ell\ell}$. The leptons from the cascade decays $\chi^{\pm}_1 \to W^*(\ell\nu)\chi^0_1,\, \chi^0_2 \to Z(\ell\ell)\chi^0_1$ are soft, limited by the inter-Higgsino splitting, while the leptons in the diboson plus jet background come predominantly from on-shell gauge bosons and are more energetic. The softness of the the signal leptons can be seen in individual lepton distributions like $p_{T,\ell}$, but it also shows up in observables constructed, such as $m_{\ell\ell}, m^{\ell\ell}_{T,2}$, etc. that are constructed from both leptons in the event. By focusing on low values for $p_{T,\ell}, m_{\ell\ell}$, etc. we can reject a large fraction of the diboson plus jet background while retaining the signal. The optimal size of the $m_{\ell\ell}$ window for capturing signal and rejecting background depends on the Higgsino mass splittings.
\end{itemize}

To show the relative size of the processes we are dealing with, in Table~\ref{table:xsecs} we show the background cross sections at various stages of the analysis, up to the final $m_{\ell\ell}$ cut. We also show the cross section for a few example parameter choices. 
\begin{table}[t!]
\begin{center}
\begin{tabular}{|c|c|c|c|c||c|c|c|c|c|}
\hline
&\multicolumn{4}{c||}{$\sigma$(fb) at 8 TeV}&\multicolumn{5}{c|}{$\sigma$(fb) at 14 TeV}\\
\hline
 &$j\ell\ell\nu\nu$ & $t\bar t$ & $j\tau\tau$ & signal &$j\ell\ell\nu\nu$ & $t\bar t$ & $j\tau\tau$ & signal & signal \\
 & & & & ($\mu = 110 \, \gev$) & & & & ($\mu = 110 \, \gev$) & ($\mu = 150 \, \gev$) \\ \hline
$p_T^j,\,  \slashed{E}_{T}>100 \, \gev$&
19.0 &   9.6 &  130.4 &    5.2 &   48.4 &   30.8 &  339.0 &   14.0  & 5.8 \\
\hline
two isolated leptons&
  17.8 &   8.8 &   46.5 &    0.8 &   45.2 &   28.0 &  120.9 &    2.2 & 0.9 \\
\hline
$m_{\tau\tau}>150 \, \gev$&
 17.3 &   8.6 &    3.7 &    0.6 &   43.9 &   27.6 &    9.7 &    1.7 & 0.7 \\
\hline
\end{tabular}
\end{center}
\caption{Cross sections (in fb) after each cut, for the major backgrounds, and the signal for $\mbox{tan}\beta = 10$, $M_1=M_2=500\,\gev$, and $\mu$ equal to $110$ or $150$~GeV. The cuts in the first row include the $b$-jet veto and a veto on events with a second jet with $p_T > 30\, \gev$. Also, the $\bar t t$ cross section has been scaled by a factor of $0.2$, the probability to fail to tag a $b$-jet.} 
\label{table:xsecs}
\end{table}
The cross sections used in Table~\ref{table:xsecs} and in all plots are the leading order (LO) values. The signal and the $WW+j$ background are initiated by the same partonic subprocess and have similar production kinematics (as $\mu \sim m_W$), so we expect the higher order corrections for these processes ($K$-factors) to be nearly identical~\cite{Fox:2012ru}\footnote{Next-to-leading-order (NLO) diboson plus jet for non-VBF topologies is currently only known to leading order.}. Including higher-order corrections would therefore increase the $S/\sqrt{B}$ by $\sim\sqrt K$. However, our analysis has also neglected several experimental details, such as lepton momentum smearing. Therefore, in an effort to compensate for the crudeness of our analysis and present a conservative result, we use the leading order cross sections\footnote{The $K$-factor for $\bar t t$ is $O(2)$~\cite{Czakon:2013goa} and potentially larger than the signal or $WW+j$. However given that it is a minor background (and the only one subject to $b$-veto and related uncertainties), we retain the leading order cross section.}. 

\subsection{Results}
\label{sec:res}

Having outlined the search strategy and identified the important backgrounds, we are ready to present our results.  We break up our search for quasi-degenerate Higgsinos into three regions, corresponding to the three parameter sets discussed in Section \ref{sec:params},
\begin{enumerate}
\item{Case I: $M_1 \gg M_2 > |\mu|$.}
We fix $\mu = 110\,\gev$, $M_1 = 1\,\tev$, and vary $M_2$ from $150\,\gev$ to $1\,\tev$.
\item{Case II: $M_2 \gg M_1 > |\mu|$.}
We fix $\mu = 110\,\gev$, $M_2 = 1\,\tev$, and vary $M_1$ from $150\,\gev$ to $1\,\tev$.
\item{Case III: $M_1 \sim M_2 > \mu$.}
We fix $M_1=M_2=500\, \gev$ and vary $\mu$ from $110\, \gev$ to $200\,\gev$. 
\end{enumerate}
The $\mu$ value is fixed to $110\, \gev$  for the first two cases to ensure that there is ample parameter space safe from the LEP II bound, and we will assume that $\mu > 0$. Unless otherwise stated, we assume $\tan{\beta} = 10$. From Eq.~(\ref{eq:dmeq1}),\,(\ref{eq:dmeq2}) we see that all splittings scale as $(1 \pm s_{2\beta})$ or $(1 \mp s_{2\beta})$ which asymptote to $1$ for large $\tan{\beta}$. Finally, the effect of raising $|\mu|$ is captured by case III.

\begin{figure}[t!]
\centering
\includegraphics[width=0.49\textwidth]{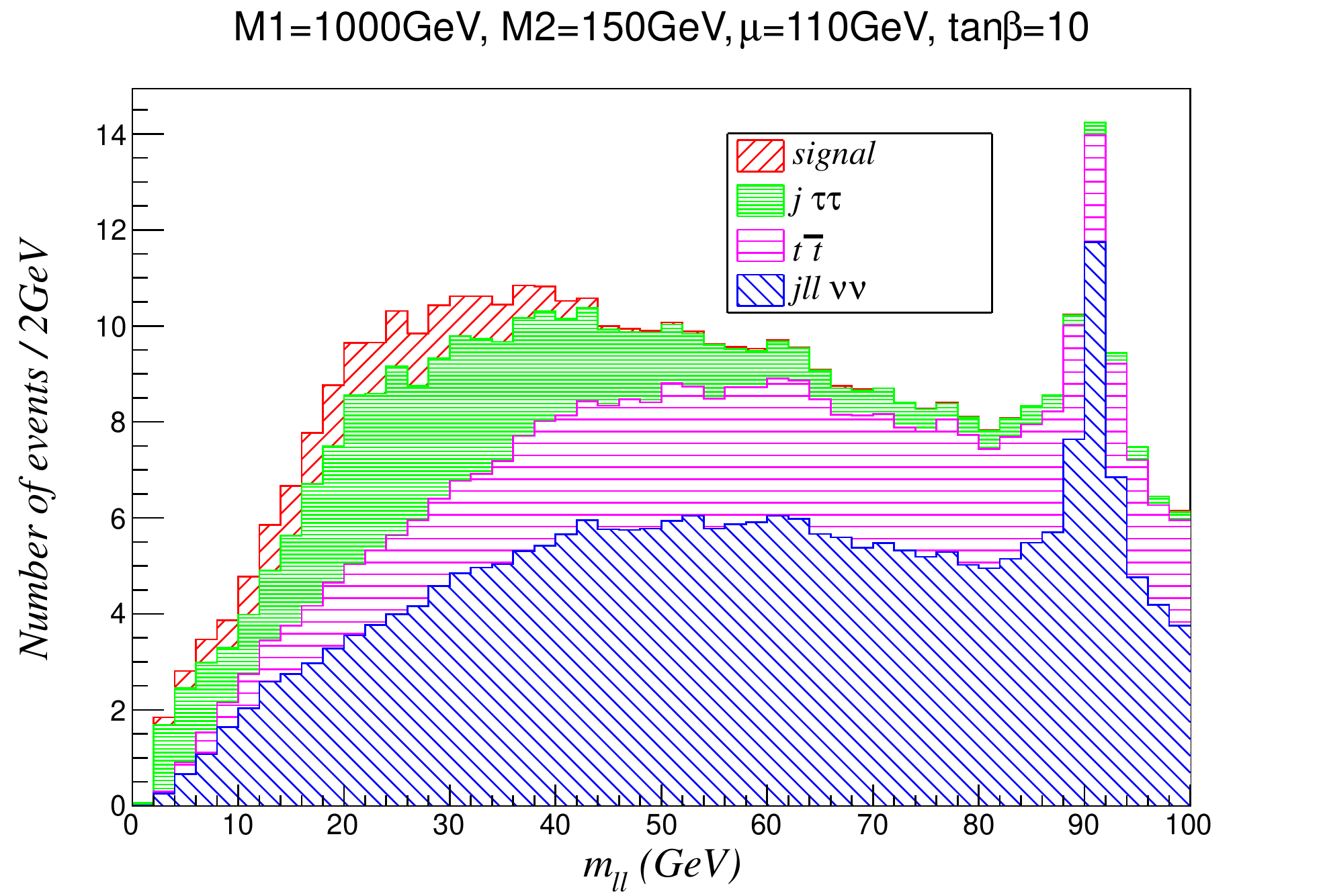}
\includegraphics[width=0.49\textwidth]{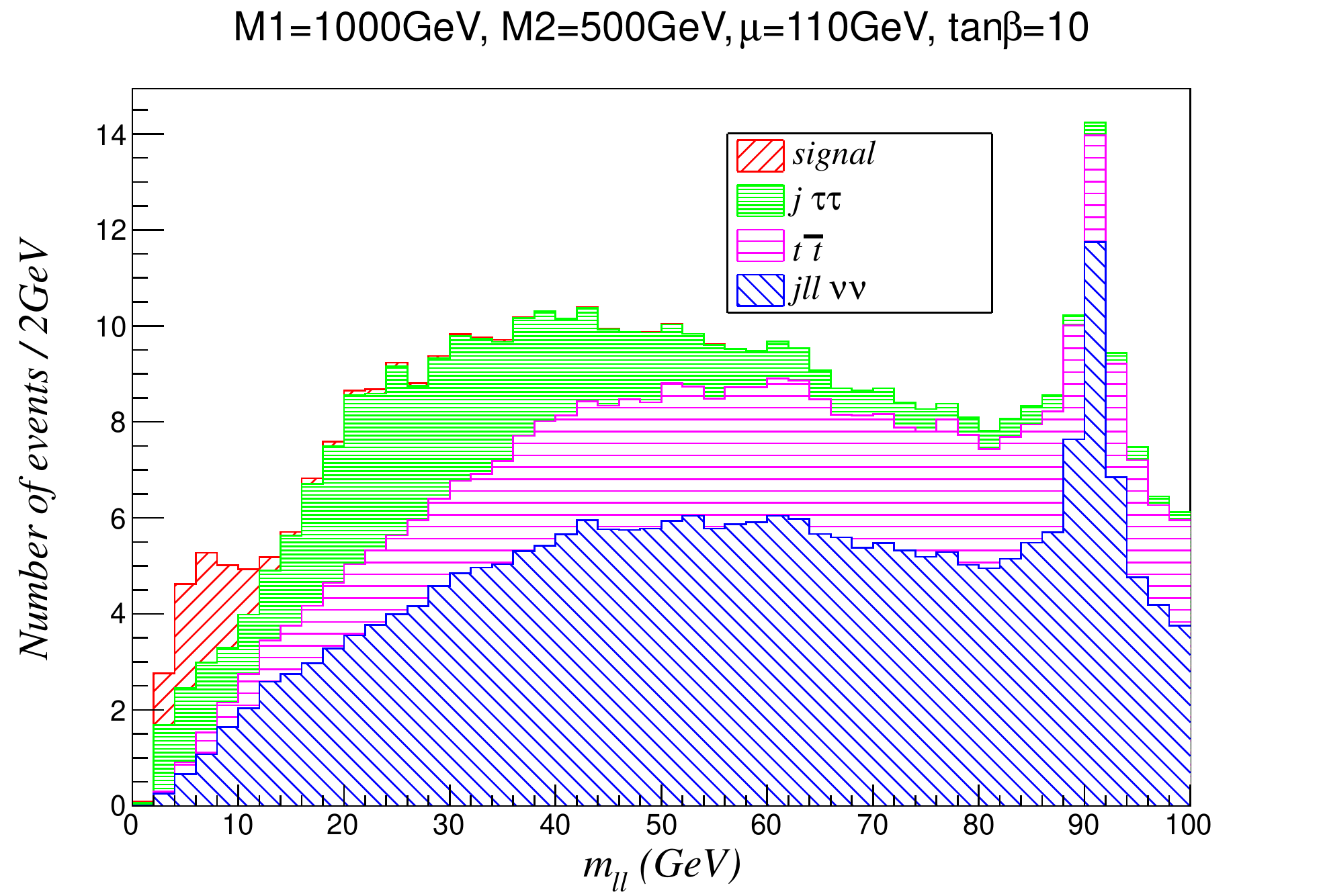}
\caption{Example $m_{\ell\ell}$ distributions after all other cuts, for case I, 20 fb${}^{-1}$ at the 8 TeV LHC\@.  Note that the parameter choices in the left side figure result in a chargino that is slightly lighter than the LEP II bound, which we show to illustrate the change in the $m_{\ell\ell}$ distribution as $M_2$ is lowered.}
\label{fig:mll_case1}
\end{figure}

We start from case I, where $M_2$ is allowed to vary with $M_1$ fixed to $1\, \tev$. Depending on the $M_2$ value,  there can be a sizable mass gap between the Higgsino states. A larger inter-Higgsino splitting means the leptons in the final state are more energetic so the efficiency (and therefore rate) for the signal events is higher. However, more energetic leptons also increase $m_{\ell\ell}$, making this distribution more similar to the background. Fixing $\mu = 110\, \gev, M_1 = 1\, \tev$ the stacked $m_{\ell\ell}$ distributions for two different $M_2$ values are shown in Fig.~\ref{fig:mll_case1}. As we will show in detail in Sec.~\ref{sec:fakes}, the fake backgrounds are $\lesssim 10\%$ of the diboson plus jet background and have similar shape.  We expect the background uncertainties (both theoretical and experimental) are at least at the 10\% level,  therefore we will neglect the fake lepton contribution in Fig.~\ref{fig:mll_case1} and in all subsequent $m_{\ell\ell}$ plots.
The $m_{\ell\ell}$ distribution for the signal in the left-hand panel is clearly similar to the background, while in the right-hand panel the signal $m_{\ell\ell}$ is clumped at lower values. The splittings in corresponding scenarios are $m_{\chi^{\pm}_1} - m_{\chi^0_1} = 5.4\,\gev, m_{\chi^{0}_2} - m_{\chi^0_1} = 51.1\,\gev  $ for left hand plot and $m_{\chi^{\pm}_1} - m_{\chi^0_1} = 5.7\,\gev, m_{\chi^{0}_2} - m_{\chi^0_1} = 21.1\,\gev $ in the right-hand plot. Note that when the two leptons in the final state come from different Higgsino decays the dilepton mass is no longer bounded by the inter-Higgsino splitting. This explains the tail of signal events in the right hand panel of Fig.~\ref{fig:mll_case1} that stretches out to $\sim40\,\gev$. 

 The other prominent feature of the $m_{\ell\ell}$ plot is the peak in the diboson background at $m_{\ell\ell} = m_Z$, which comes from $pp \to Z(\nu\bar{\nu})Z(\ell\ell) + j$. This feature will be cut out once we select a $m_{\ell\ell}$ window to determine the final significance, but in practice it may be useful as a control sample, i.e.  to pin down the diboson plus jet background normalization. 
 
 The $m_{\ell\ell}$ distributions at $8\, \tev$ and $14\, \tev$ look almost identical, as we will show explicitly later on. This is easy to understand; Higgsino production, diboson production, and Drell-Yan $\gamma^*/Z$ all require a quark-antiquark initial state. Once we ask for an additional jet, the dominant partonic subprocess (at the LHC) for all three of these process is gluon plus quark, so the change in parton luminosity moving from 8 TeV to 14 TeV will affect all three processes in the same way. The scale of the signal is slightly different than the background, since $2\,m_{\chi} \sim 2\,\mu > 2\, m_W > m_Z$. However, all of these scales are small compared to the beam energy, so the difference between the signal and background scales has negligible effect. The $\bar t t$ background is primarily initiated by gluon-gluon collisions, so it will rescale slightly differently as the collider energy is changed.

Moving to case II, $M_1$ varies while $M_2 = 1\,\tev$ is fixed. We show the $m_{\ell\ell}$ distributions for two sample $M_1$ values below in Fig.~\ref{fig:mll_case2}, with $\mu = 110\, \gev, M_2 = 1\, \tev, \tan{\beta} = 10$. For the two cases, the splittings are $m_{\chi^{\pm}_1} - m_{\chi^0_1} = 18.8\,\gev, m_{\chi^{0}_2} - m_{\chi^0_1} = 25.9\,\gev $ for the left spectrum, and $m_{\chi^{\pm}_1} - m_{\chi^0_1} = 6.4\,\gev, m_{\chi^{0}_2} - m_{\chi^0_1} = 12.3\,\gev $ for the right. As in case I, there is a clear separation between signal and background when $m_{\ell\ell}$ is small (right-hand panel), but this distinction evaporates as the splitting increases (left-hand panel).  The difference in the signal $m_{\ell\ell}$ spectrum between case II and case I can be traced to the $\chi^0_2$ branching ratios shown in Fig.~\ref{figure:then2BR}. When $M_1$ is light, as in case II, $\chi^0_2$ decays predominantly to $Z^*$. The two leptons in the case come from a common mother particle, so the $m_{\ell\ell}$ in this case is bounded above by the mass of the $Z^*$. For the right-hand spectrum in Fig.~\ref{fig:mll_case2} this maximum value is $\sim 12\, \gev$, and indeed we see that the majority signal events fall below this value. In case I, $M_2$ is light and $\chi^0_2 \to W^{*\pm}\chi^{\mp}_1$ dominates, so the two leptons in the final state often come from two different particles; either each Higgsino in the event gives a lepton, or the two leptons come from two successive decays of $\chi^0_2$, $\chi^0_2 \to W^{*\pm}(\ell\nu)\chi_1^{\mp}, \chi^{\mp}_1 \to W^{*\mp}(\ell\nu)\chi^0_1$. Either way, the $m_{\ell\ell}$ distribution is more spread out, since the dilepton mass is not fixed to be below the inter-Higgsino splitting. 
\begin{figure}[t!]
\centering
\includegraphics[width=0.49\textwidth]{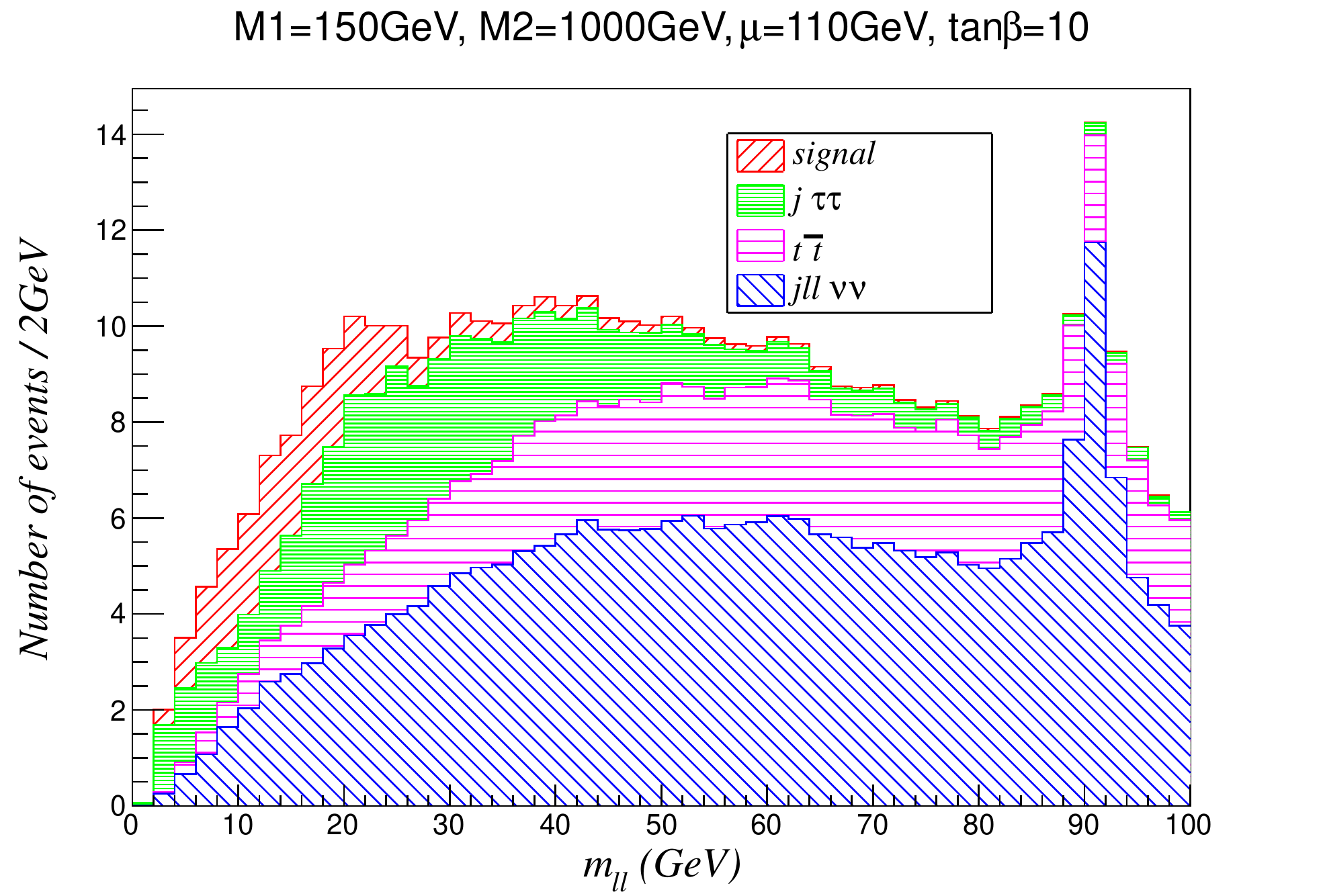}
\includegraphics[width=0.49\textwidth]{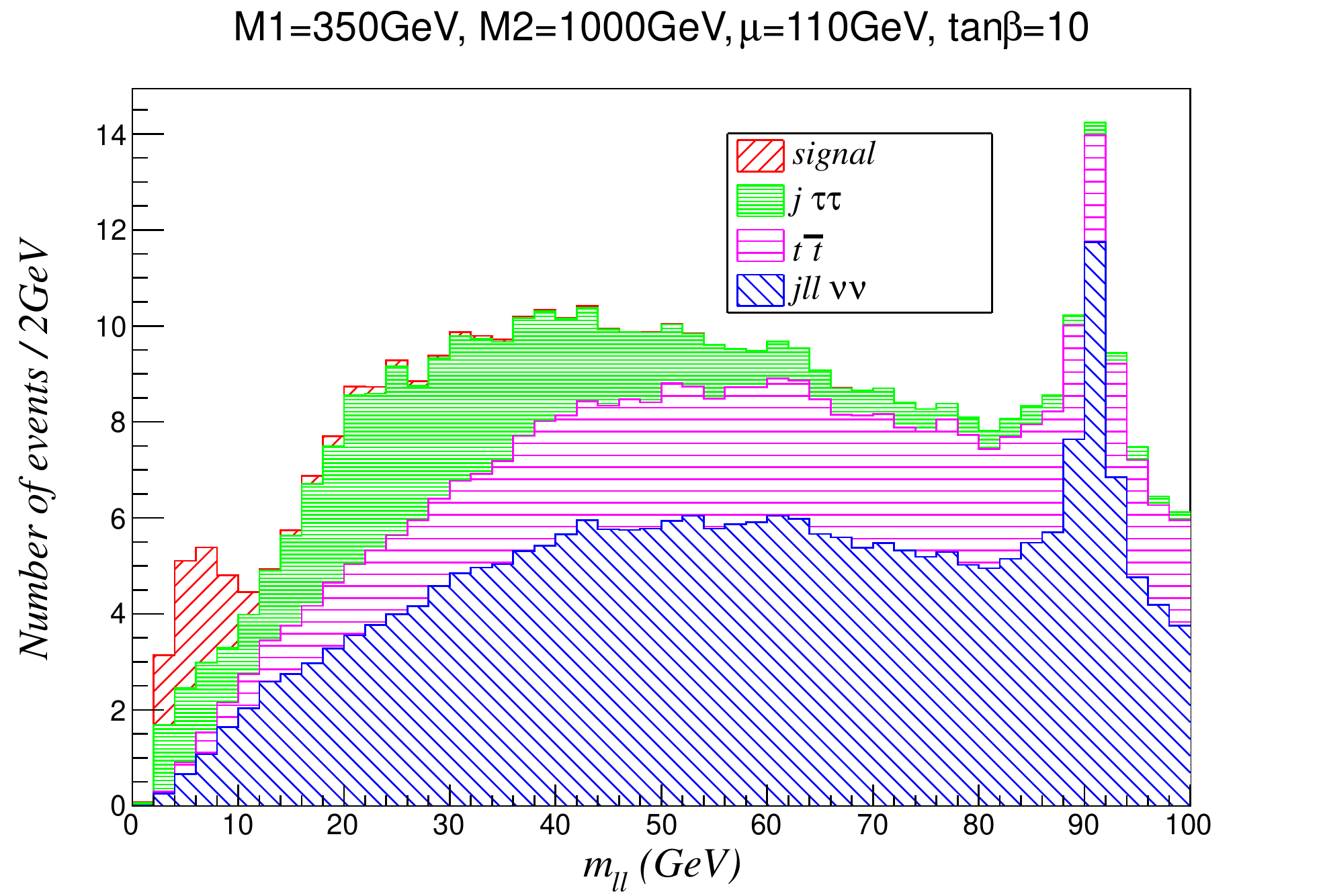}
\caption{Example $m_{\ell\ell}$ distributions after all other cuts, for case II, 20 fb${}^{-1}$ at the 8 TeV LHC.}
\label{fig:mll_case2}
\end{figure}

The difference between case I and II brings up a question about the signal, namely the breakdown between same-flavor and unlike-flavor events. If both leptons in the event come from a single $Z^*$ then they will have the same flavor, while if the leptons come from two different decays, either two separate $\chi^{\pm} \to W^* \chi^0_1$ decays, a single $\chi^0_2 \to W^* \chi^{\pm}_1 \to W^*W^*\chi^0_1$ decay, or some other combination, they will only have the same flavor $50\%$ of the time. The percentage of the signal coming from same-flavor decays for $\mu = 110\, \gev,\, \tan{\beta} = 10$ is shown below in Fig.~\ref{fig:SFfraction}. The fraction is shown for a range of $M_1, M_2$ and includes all cuts except the final $m_{\ell\ell}$ window -- this is important to remember since the efficiency for an event with two leptons from a single $Z^*$ is generally not the same as the efficiency for an event where the leptons come from two different decays.
\begin{figure}[t!]
\centering
\includegraphics[width=0.45\textwidth]{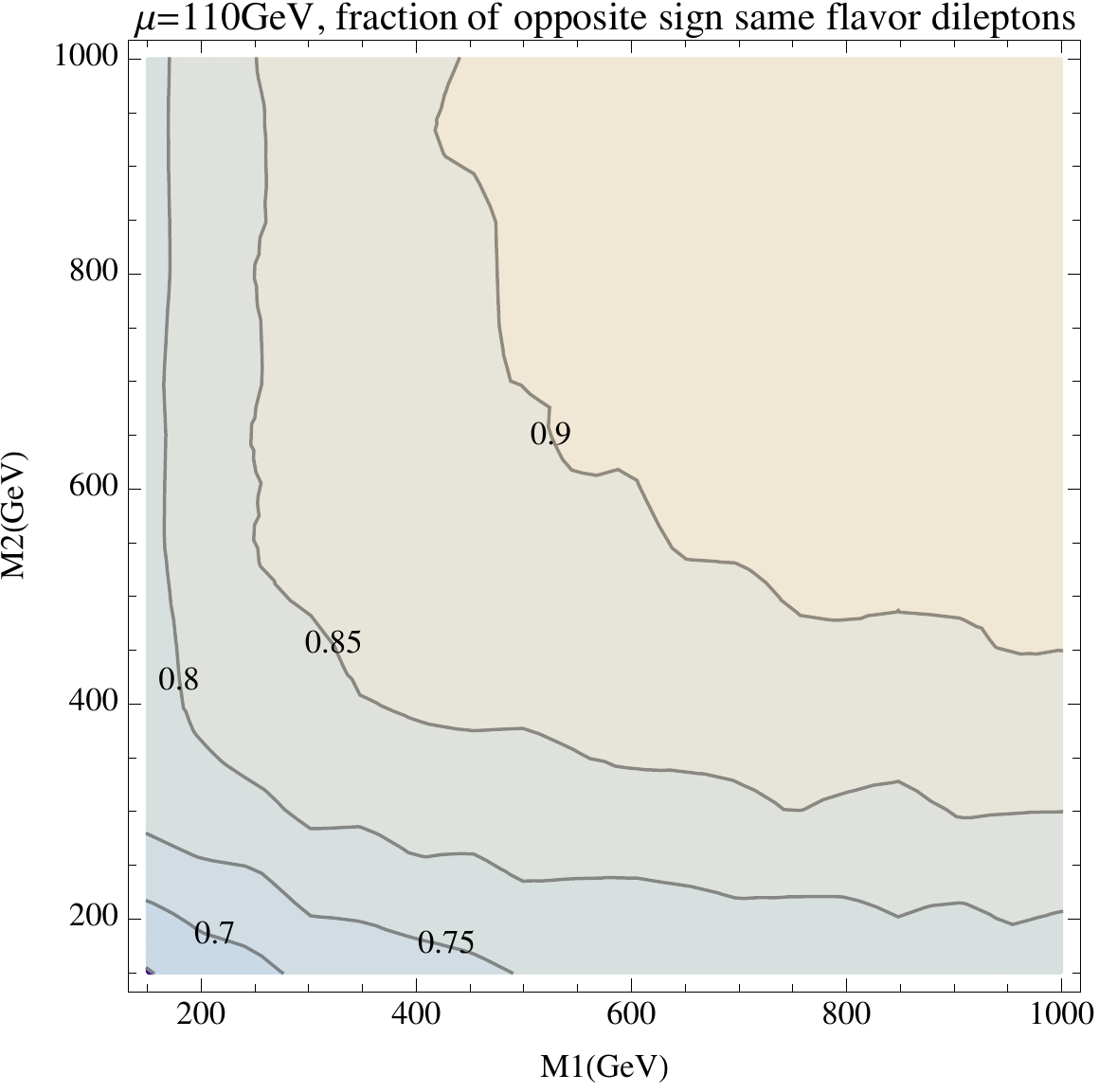}
\caption{Contours of the fraction of the signal events $j + \slashed E_T +  \ell\ell$  in cases I and II with an opposite-sign, same-flavor lepton pair after all analysis cuts (except the $m_{\ell\ell}$ window cut) have been applied.  In this figure $\mu = 110$~GeV for LHC operating at $8$~TeV, though the contour fractions are very similar for larger $\mu$ as well as for LHC operating at $14$ TeV, considered elsewhere in the paper.  Note that for this particular value of $\mu$, the region below 
$M_2 \lesssim 450$~GeV is excluded by LEP II.}
\label{fig:SFfraction}
\end{figure}
In principle this lepton-flavor information could be used to further diversify and optimize our search. For example, the same-flavor fraction of the $WW+j$ and $\bar t t$ backgrounds (pre-cuts) is only 50\%, significantly less than the signal. Since the $Z/\gamma^* +j $ comes from $\tau$ decays, it will also have a nearly 50\% same-flavor fraction. The only background with a $\sim 100\%$ same-flavor lepton fraction is the QCD resonance background that we will discuss in the next section. Therefore, by requiring unlike-flavor leptons, we can create a QCD-free and nearly signal-free control region that can be used to verify background modeling. At the same time, the same-flavor fraction is more signal-rich, but suffers from smaller statistics and potential low-energy QCD backgrounds, though as we will demonstrate in Sec.~\ref{sec:fakes} we believe these to be small for our analysis. Furthermore, breaking the signal down further into individual lepton flavors may also be useful given that QCD fake-leptons are much more likely to be electrons. We have not attempted a detailed signal breakdown here, but this is an interesting direction to pursue more thoroughly in the future.

Having summarized the key distributions and properties of the signal, we now want to turn these distributions into an estimated significance. The significance is determined by taking a $m_{\ell\ell}$ window and counting the number of signal and background events inside. The size of the window is chosen to optimize $S/\sqrt B$ for the given set of inputs $\mu, M_1, M_2, \tan{\beta}$. For scenarios with relatively small splitting, such as the right hand panels of Fig.~\ref{fig:mll_case1},\,\ref{fig:mll_case2}, the optimal window size is small. As the Higgsino mass splitting grows and the $m_{\ell\ell}$ spreads out, we are forced to take larger windows that let in more background. The resulting significance for cases I and II are shown below in Fig.~\ref{fig:significance_case12}. We show the significance for two different scenarios: i.) using the current data set,  $20\, \fbinv$ at 8 TeV, and ii.) assuming $100\, \fbinv$ of 14 TeV LHC running. When calculating the significance we require at least 5 signal events in the $m_{\ell\ell}$ window\footnote{For small numbers of events, the uncertainties are not quite Gaussian. We ignore this difference and stick with $S/\sqrt B$ as a rough measure of significance.}. 
\begin{figure*}[t!]
\centering
\begin{tabular}{cc}
\includegraphics[width=0.5\textwidth]{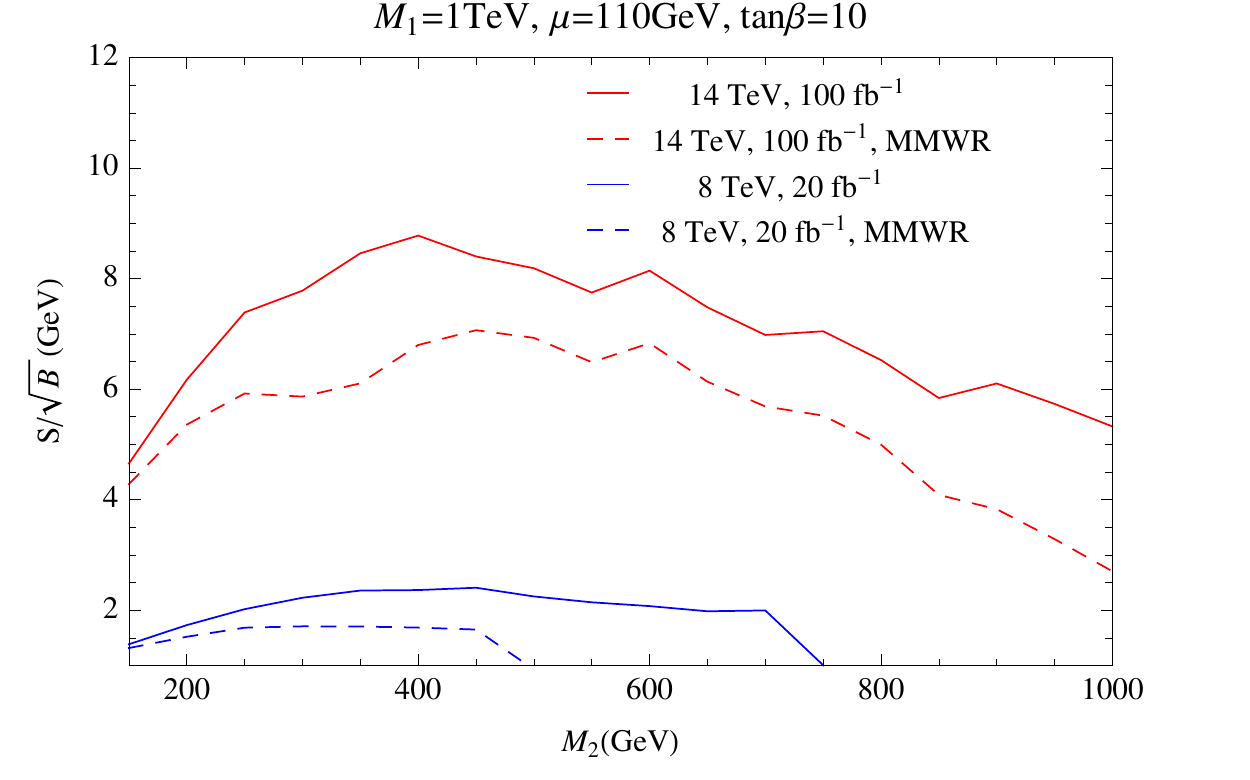}
\includegraphics[width=0.5\textwidth]{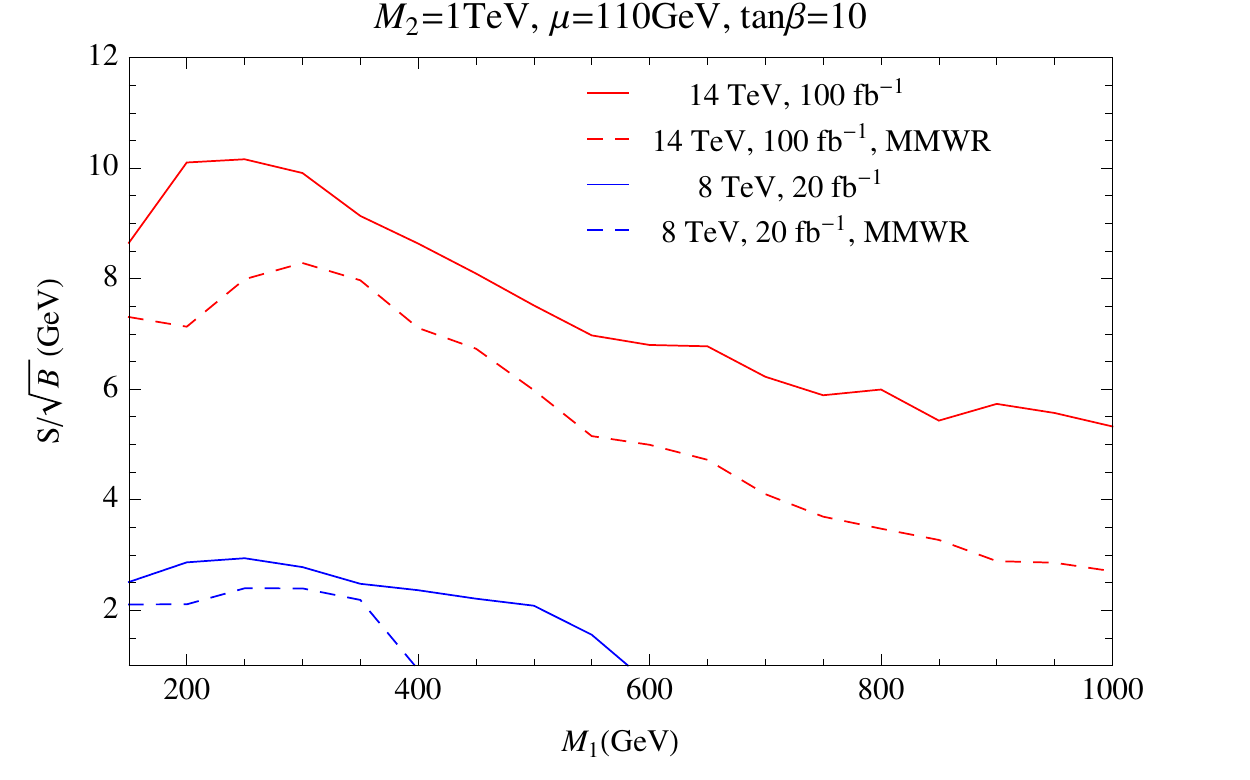}
\end{tabular}
\caption{The significance for case I and case II. The dashed lines show the significance if the QCD resonance region of $m_{\ell\ell}$ is removed. See below for details.}
\label{fig:significance_case12}
\end{figure*}
We will argue in Sec.~\ref{sec:fakes} that the background containing low-mass QCD resonances is small, however one way to reduce this background even further is to veto the $m_{\ell\ell}$ events in the region of J/$\psi$ and below, $m_{\ell\ell} < 4\,\gev$, as well as the $\Upsilon$ region, $8\,\gev < m_{\ell\ell}< 12\,\gev$.\footnote{Since events were binned in $2$~GeV bin sizes, we did not attempt to optimize the MMWR region further.} The significance including this cut is labeled as ``MMWR'' in Fig.~\ref{fig:significance_case12}, which stands for ``meson mass window removed''. \\

\begin{figure*}[t!]
\centering
\begin{tabular}{cc}
\includegraphics[width=0.5\textwidth]{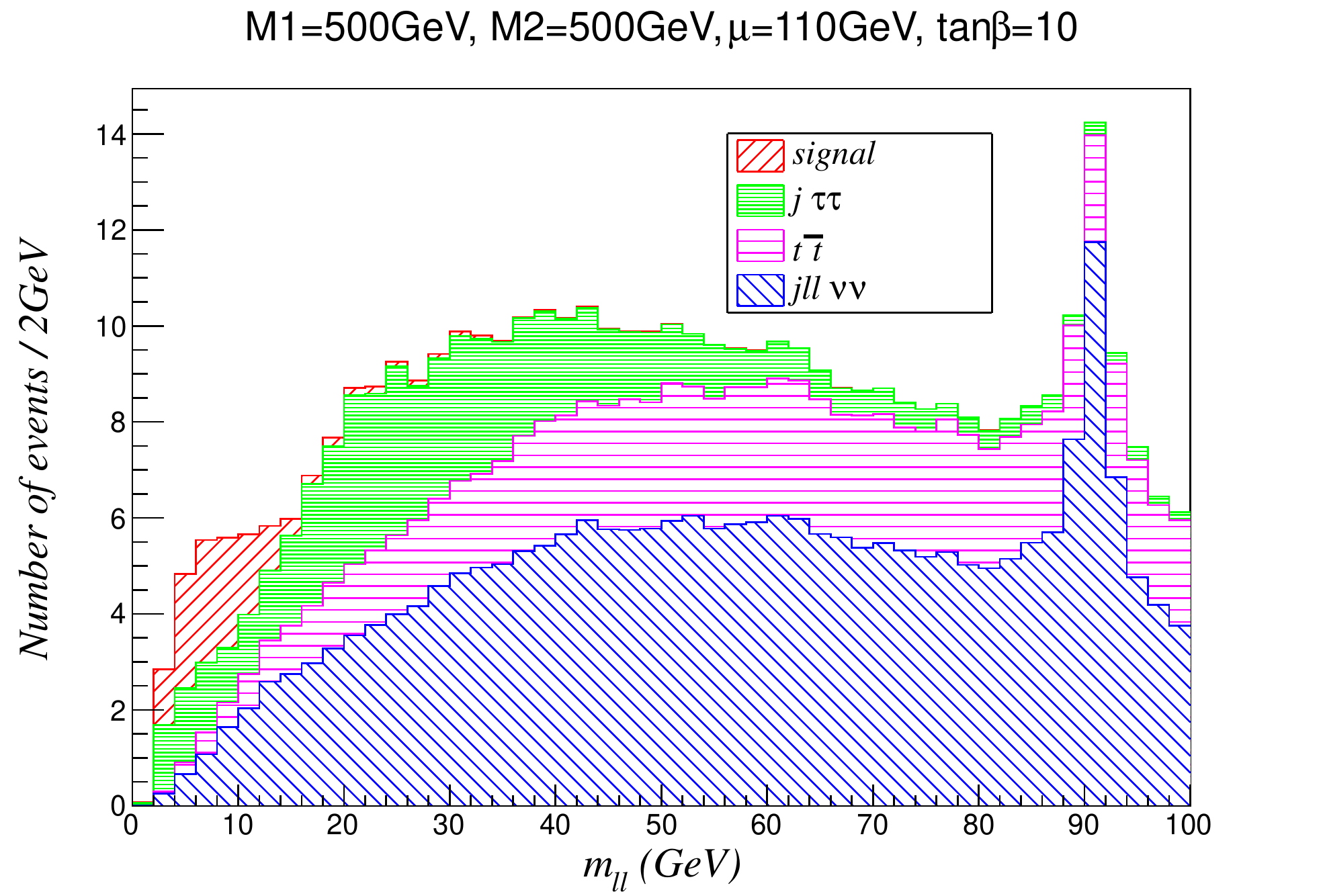}
&\includegraphics[width=0.5\textwidth]{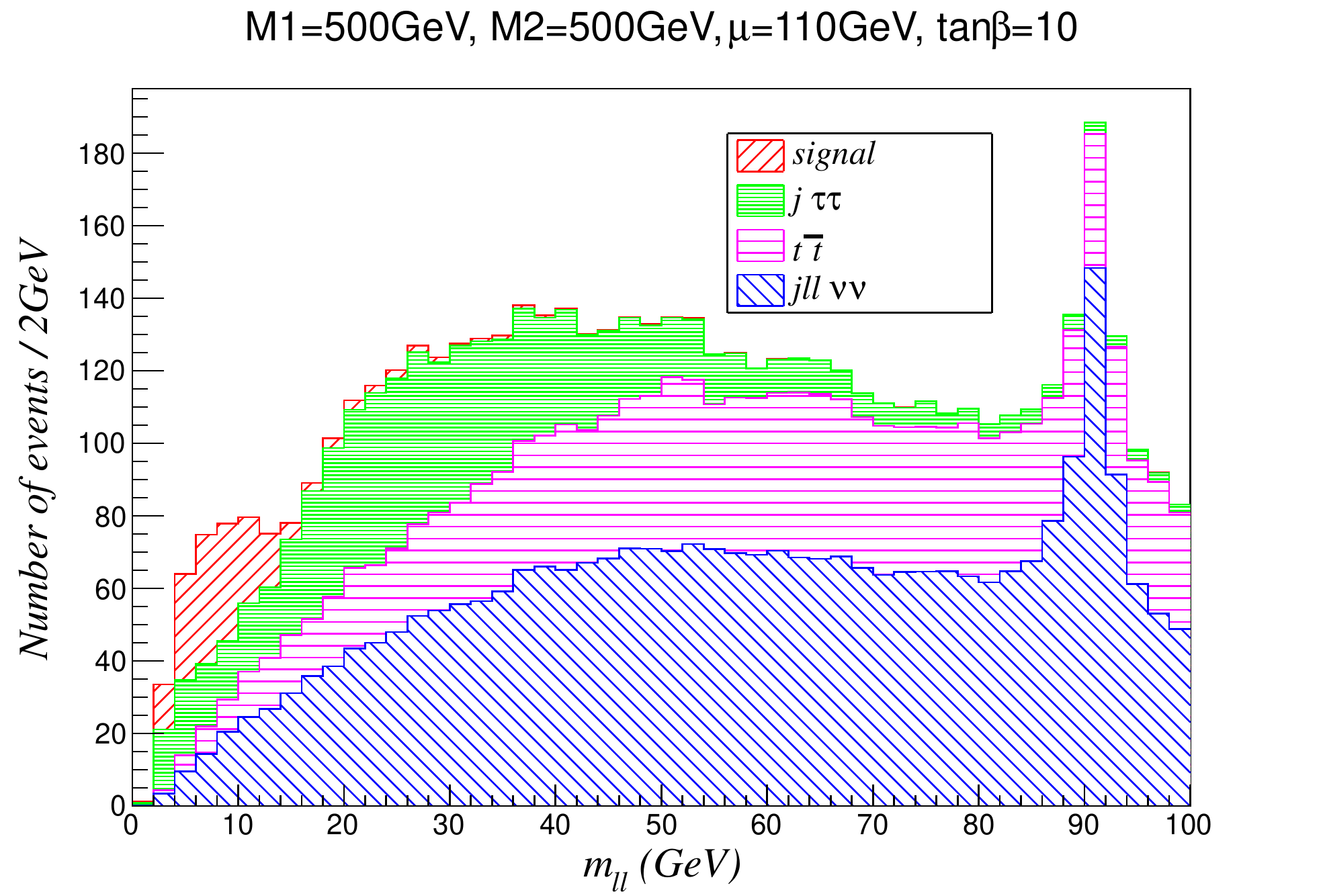}
\end{tabular}
\caption{The $m_{\ell\ell}$ distributions (stacked) after all other cuts, for $M_1=M_2=500\,\gev$ and $\mu = 110\,\gev$. Left: 20 fb${}^{-1}$ at LHC, $\sqrt s = 8\,\tev$; right: 100 fb${}^{-1}$ at the $\sqrt s = 14\, \tev$ LHC.}
\label{fig:mll_case3}
\end{figure*}

Finally, we move to case III, where $M_1,\, M_2$ are fixed to $500\, \gev$ and $\mu$ is allowed to vary. In Fig.~\ref{fig:mll_case3}, we show the $m_{\ell\ell}$ distributions after all other cuts have been applied for $\mu=110\,\gev$. The left-hand panel of Fig.~\ref{fig:mll_case3} shows the  $m_{\ell\ell}$ distribution at a $8\,\tev$ center of mass LHC, while the same distribution for a $14\, \tev$ collider is shown in the right-hand panel; the shapes  of the distributions at the two different energies are nearly identical, as expected from earlier arguments. Interestingly, in this case, almost all signal events are concentrated in a few low $m_{\ell\ell}$ bins, where the backgrounds are relatively small. Therefore, it is clear that we should cut off all events with higher $m_{\ell\ell}$ to maximize the significance. Due to the low efficiency for the leptons to pass the momentum threshold, the event rate is below 5 events at the $\sqrt s = 8\,\tev$ LHC with $20$~fb$^{-1}$ of data when $\mu \gsim 140\,\gev$. On the other hand, at the $\sqrt s =14\,\tev$ LHC, the statistics is good enough to reach a $\sim 5\,\sigma$ discovery with a good $S/B$ ratio when $\mu\lsim135\,\gev$. For example, for $\mu=120\,\gev$, choosing the mass window as $0<m_{\ell\ell}<12\,\gev$, we obtain 103 signal events and 197 background events for 100 fb${}^{-1}$, yielding $S/\sqrt{B}=7.3$.  

\begin{figure}[t!]
\centering
\includegraphics[width=0.65\textwidth]{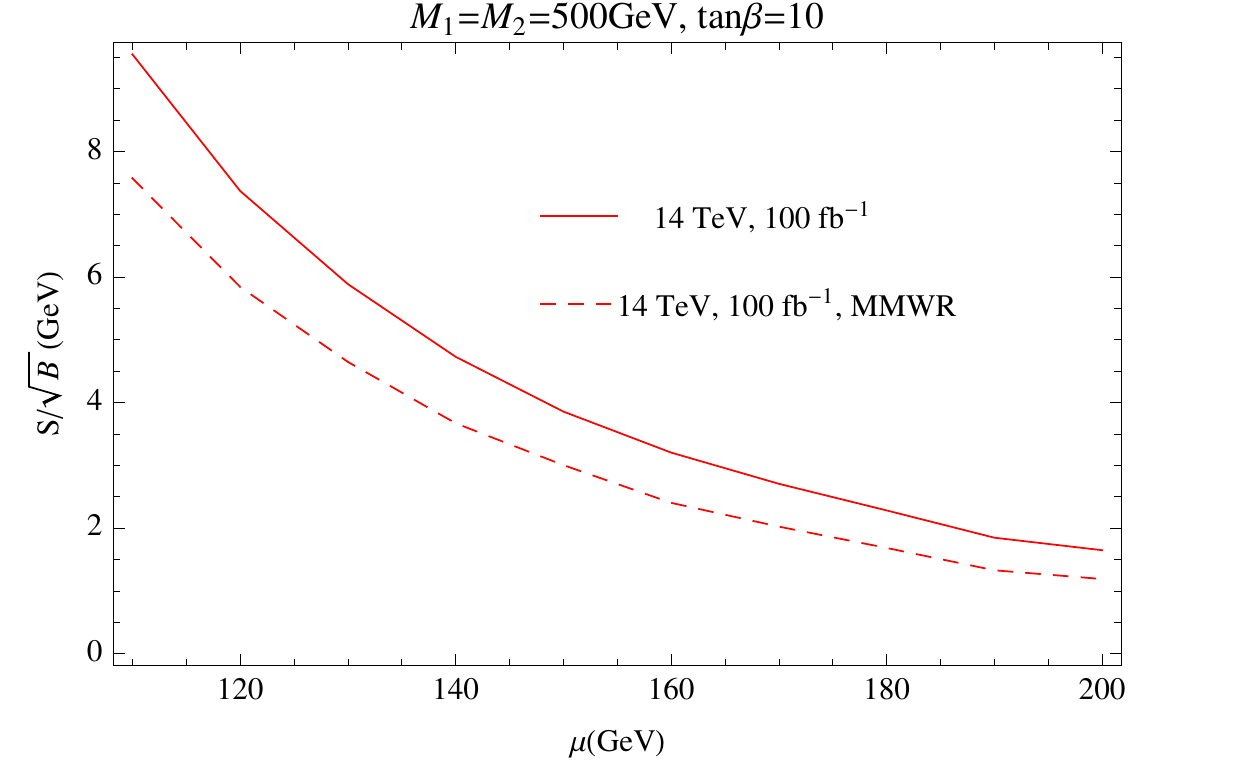}
\caption{Significance for case III. As in Fig,~\ref{fig:significance_case12}, the dashed lines show the significance if the QCD resonance region of $m_{\ell\ell}$ is removed.}
\label{fig:significance_case3}
\end{figure}

When $M_1$ and $M_2$ are held at $500\,\gev$ and $100 < \mu < 200$~GeV, the mass splittings remain almost unchanged. Therefore the above discussion applies for the whole region. However, the event rate drops rapidly when $\mu$ is increased. The $S/\sqrt{B}$ values for this case are shown in Fig.~\ref{fig:significance_case3} below assuming 100 fb${}^{-1}$ at the $\sqrt s = 14\,\tev$ LHC. 

\begin{figure*}[t!]
\centering
\begin{tabular}{cc}
\includegraphics[width=0.5\textwidth]{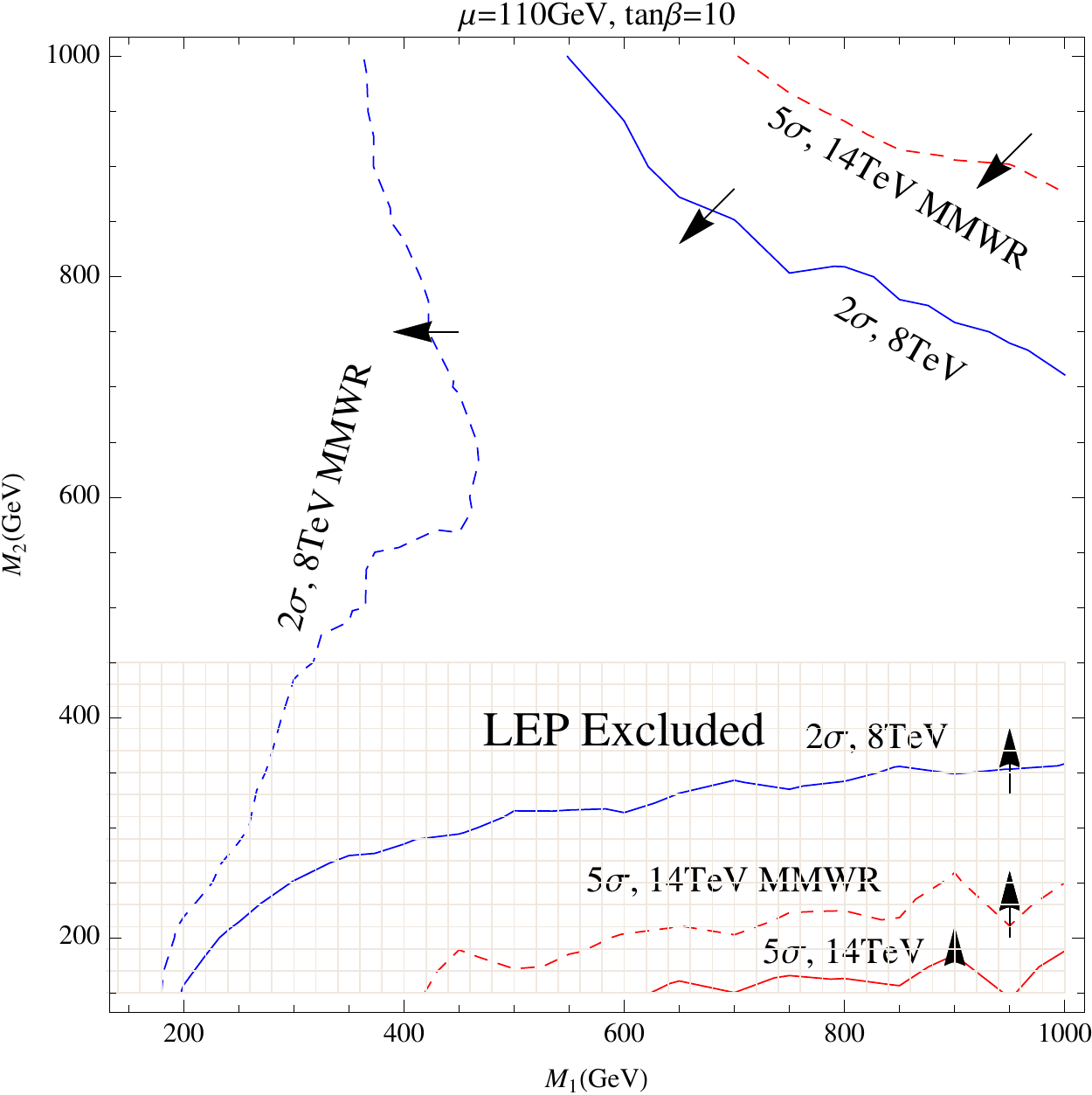}
&\includegraphics[width=0.5\textwidth]{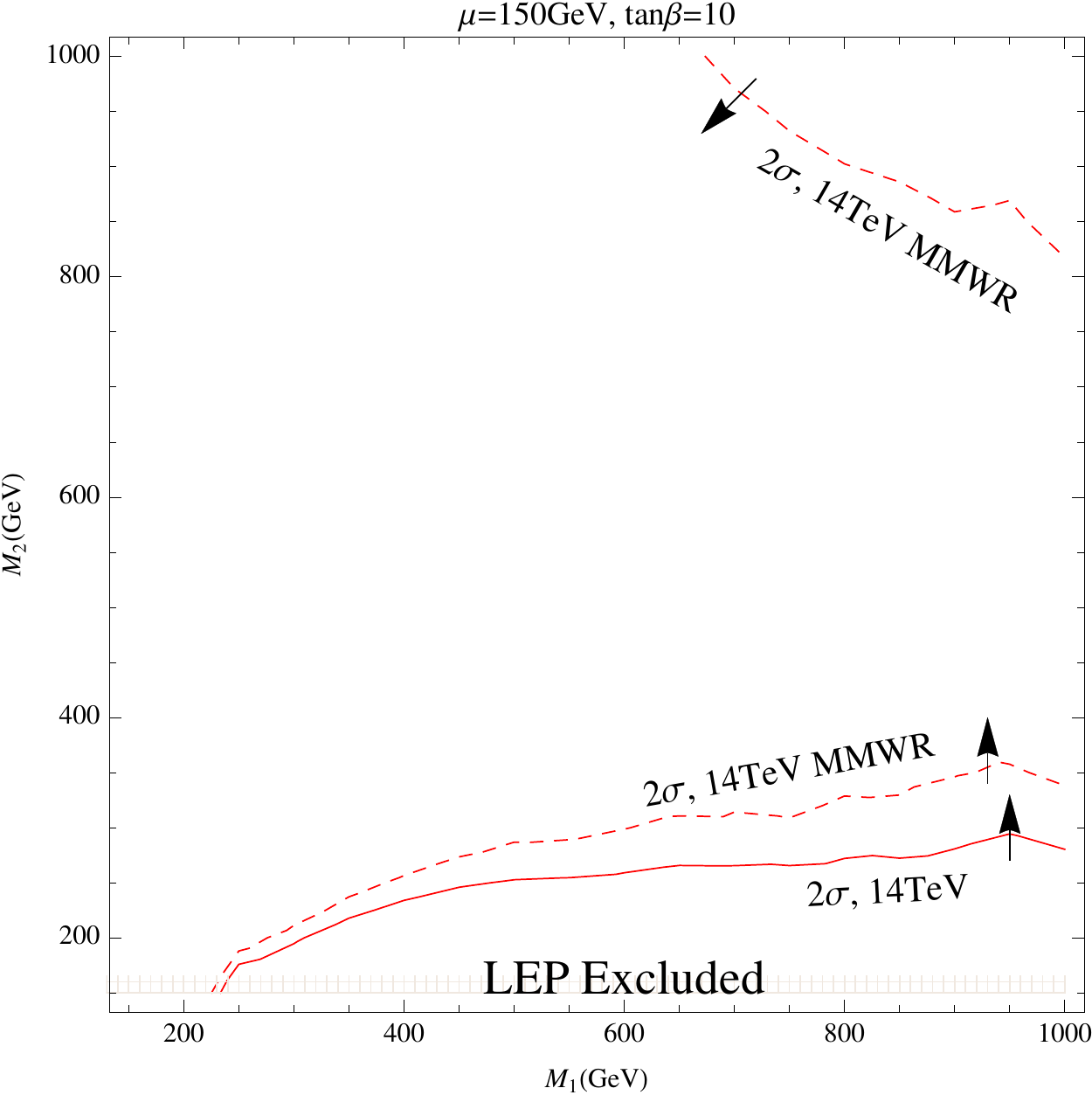}
\end{tabular}
\caption{The LHC discovery potential for two slices of the $(\mu,M_1,M_2)$ parameter space. Left: $\mu=110\,\gev$, 2 $\sigma$ contour for 20 fb${}^{-1}$ at the $8\,\tev$ and 5 $\sigma$ contour for 100 fb${}^{-1}$ at $14\,\tev$; right: $\mu=150\,\gev$,  $2\,\sigma$ contour for 100 fb${}^{-1}$ at $14\,\tev$. As in Fig,~\ref{fig:significance_case12}, the dashed lines show the significance if the QCD resonance region of $m_{\ell\ell}$ is removed.  Higher significance is obtained by moving in the direction of the arrows shown on the figures.}
\label{fig:reach}
\end{figure*}

We summarize the reach of the LHC for $\mu=110\, \gev$ and $\mu=150\,\gev$ in Fig.~\ref{fig:reach}. We see that for $\mu=110\, \gev$, the $8\,\tev$ LHC can only probe the signal at 2\,$\sigma$, while the $14\,\tev$ LHC can discover the signal at the $5\,\sigma$ level on the entire $(M_1, M_2)$ plane, except for the low $M_2$ region which has already been ruled out by LEP. For $\mu=150\,\gev$, we have $2\,\sigma$ sensitivity for most part of the parameter space with 100 fb${}^{-1}$, which of course can be improved with larger integrated luminosity. 

\subsection{Additional backgrounds: fake leptons and double-parton scattering}
\label{sec:fakes}

While the dominant backgrounds were identified in the previous section, here we consider other processes coming from the reality of the LHC environment, i.e.\ the limited ability to distinguish different types of particles and the fact that collisions occur amidst a soup of multiple soft-QCD interactions. Given that our signal cross section is quite small and contains relatively low-$p_T$ objects, it is especially important to verify these ``environmental'' backgrounds are under control.  All cross sections in this section are calculated at leading order using Madgraph 5 at $\sqrt{s} = 8$~TeV using the default parton-distribution functions and scale choices. These numbers should be taken as rough estimates given that environmental backgrounds come with multiple sources of uncertainty. 

One frequent source of environmental background is jets that fake leptons. We performed two studies to estimate the background from jets faking leptons; one study focused on light-flavor jets (and gluons) faking leptons, and a separate study on heavy flavor jets. A separate, heavy-flavor study is motivated because $b/c$ quarks can promptly decay to leptons and are therefore far more likely to fake an isolated lepton than light jets are. The most likely way fake leptons can mimic our final state is from $W(\ell \nu) + $jets  ($W(\ell \nu) + \bar b b, \bar c c$ for heavy flavor). In this case, the $\slashed E_T$ and one of the required leptons comes from the decay of the $W$, one of the jets is hard and central enough to satisfy our analysis cuts, and the second (or higher multiplicity) jet in the event fakes a lepton.

 Focusing on the $W+$ light-flavor jet case first, the starting point is the cross section for $W(\ell\nu) + j$ after requiring exactly one hard jet ($p_{T,j} > 100\, \gev, |\eta_j| < 2.5$) and $\slashed E_T > 100\, \gev$. At 8 TeV, we find $\sigma(p p \to W(\ell \nu) + j)_{jet, \slashed E_T\, cuts} \simeq 38\, \pb$. We generate events with this topology, with showering/hadronization as before, and clustering the output into jets.  Extra radiation emitted during the showering will often cause the $W+j$ events to appear to contain multiple jets. The jet multiplicity breakdown of the events post-showering/hadronization depends on the jet definition, which we choose to be $p_T > 7\, \gev, |\eta_j| < 2.5$, identical to the lepton requirements.  To get an estimate of the efficiency these $W+j$ events would have should one of the jets fake a lepton, we randomly remove one of the subleading jets and call it a lepton, then feed the modified event into the rest of the analysis chain. This is only an approximation. In reality the fake rate depends on the kinematics ($p_T, \eta$) of the jet, however we will ignore this for now and assume the fake rate is constant. Treating one of the jets as a lepton, we find the cross section after all cuts (except $m_{\ell\ell}$) is $6.8\, \pb \times \epsilon$, where $\epsilon$ is the fake-rate. We can make this approximation more accurate by accounting for the fact that there is sometimes more than one soft jet around capable of faking a lepton. In our sample of $W(\ell \nu)$ + hard jet events, we find the breakdown is 60\% events with 1 candidate jet: 30\% events with 2 candidate jets: 10\% events with 3 candidate jets\footnote{To be clear: all events have one jet by construction, so this breakdown refers to how many events have 2 jets ($p_{T,j} > 7, |\eta_j| < 2.5$) vs. 3 jets vs 4. We find that roughly 1/3 of all $W+j$ events have at least one extra jet given the parton level cuts and jet definition used here.}. Including a factor of $\epsilon$ for every candidate jet, the modified rate becomes $6.8 \pb \times 1.5\, \epsilon = 10.2\, \pb \times \epsilon$.  To gauge the total impact of the fakes, we need a fake rate. Curtin et al~\cite{Curtin:2013zua} recently performed a detailed study containing several parameterizations of the lepton fake rates extracted from comparison with CMS data\footnote{Their numbers are also consistent with ATLAS studies such as~\cite{ATLASfakes}.}. At low $p_T$, where the bulk of our fakes lie given our lepton $p_T$ cut, they quote fake rates of $O(0.6 -3\times 10^{-5})$. Using these numbers, the background from light-jet fakes is $O(\%)$ of the dominant background. Even if we use a more conservative fake rate of $10^{-4}$ -- motivated perhaps by the low lepton threshold -- the fake background is only 6\% of $WW + j$.  Given the size of our signal, a minor O(5-10\%)-level background could still be problematic if it had a $m_{\ell\ell}$ shape similar to the signal. However, we find this is not the case here, as the $m_{\ell\ell}$ distribution from $W+j$ events with a fake lepton is shifted to higher $m_{\ell\ell}$ values than the signal.   The $W+j$ events still have one energetic lepton from the decay of the on-shell $W$, and the angular separation between the `real' and `fake' leptons is typically large\footnote{The lepton from the $W$ decay typically points away from the ISR jet, while the second `lepton' (= soft jet)  is often close to (in $\Delta R$) the ISR jet -- the result of soft/collinear-enhanced radiation that happened to be far away enough from the original jet to be classified as a separate object.}. Both features push the $m_{\ell\ell}$ distribution for the fake background out of the signal region.
 
To study the effect of fake leptons from heavy quarks, we looked at a more flavor-enriched sample, $W(\ell\nu) + \bar b b$. To estimate the fake rate here, we generated $W + \bar b b$ events and fed them through our established isolation, clustering, and analysis chain, treating the events as any other (non-environmental) background. This directly tests how often leptonic decay products contained in jets originating from $b$-jets pass the lepton isolation requirements described in Sec.~\ref{sec:limits}. At 8 TeV, we find the cross section after all cuts except $m_{\ell\ell}$ is $O(1\, \fb)$, $\sim\,5\%$ of $WW+j$. As in the study of fake leptons from light-flavor jets above, we must check that the $W(\ell\nu) + \bar b b$ fake signal has a $m_{\ell\ell}$ distribution that is distinct from the signal. The $m_{\ell\ell}$ distribution from the fake background is shown in Fig.~\ref{fig:minor_bg} below.  While the $m_{\ell\ell}$ distribution from $W(\ell\nu) + \bar b b$ fakes is not as broad as the $m_{\ell\ell}$ distribution for $WW+j$, there are few events in the signal region.
\begin{figure}[t!]
\centering
\includegraphics[width=0.65\textwidth]{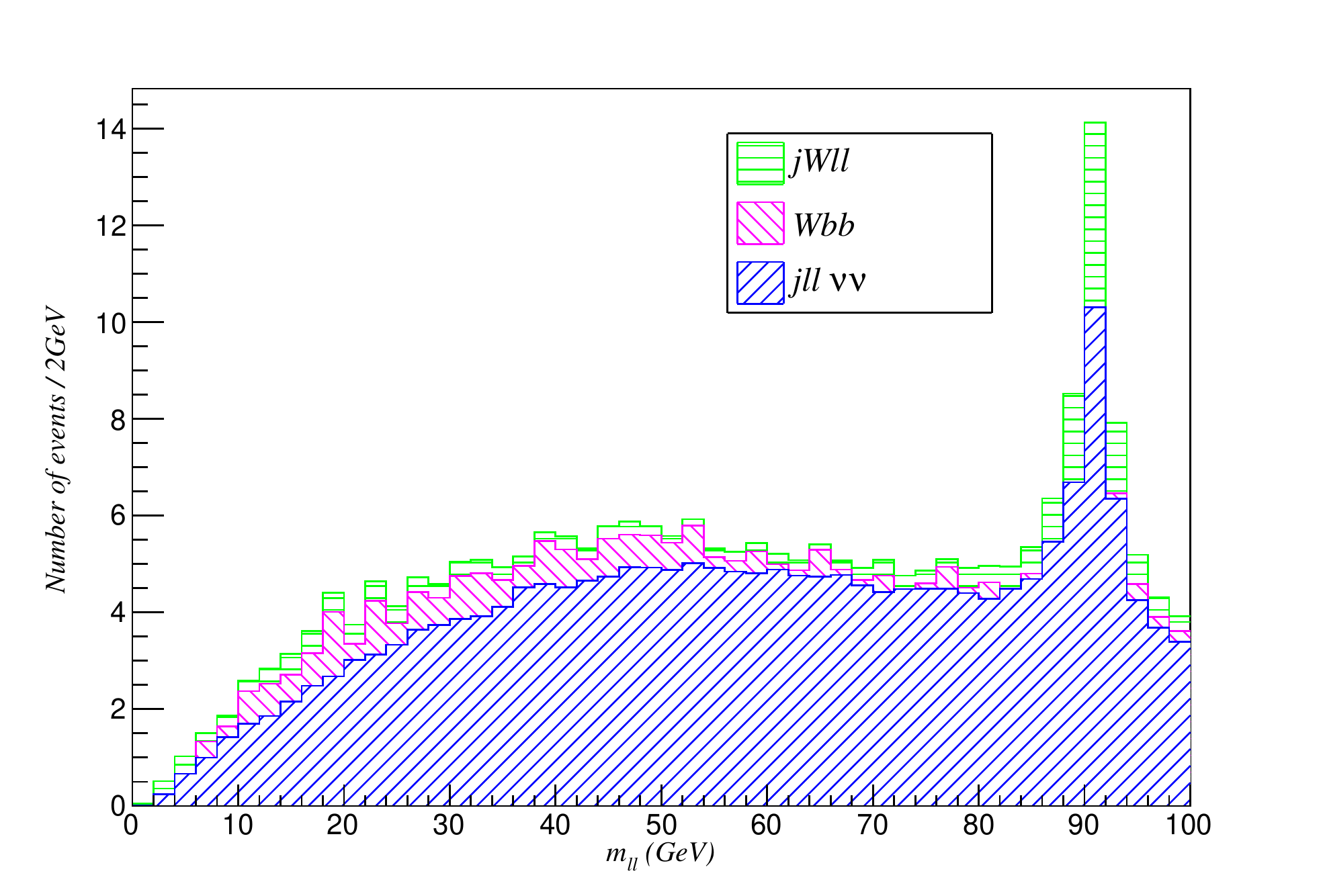}
\caption{The dilepton invariant mass distribution at the 8 TeV LHC, after all cuts, for the $W/Z + \bar bb$ fake-lepton background (pink) compared with the distribution from the diboson ($\ell\ell+X$) plus jet contribution (blue). Additionally, the $m_{\ell\ell}$ from three-lepton diboson plus jet events ($WZ/\gamma^*(3\ell+\nu) + j$) is shown in green. The $WZ/\gamma^* + j$ events can mimic our signal final state if one of the leptons is lost. All three backgrounds have similar $m_{\ell\ell}$ shape. }
\label{fig:minor_bg}
\end{figure}

Another environmental background we must consider is double-parton-scattering (DPS), two hard collisions within the same colliding protons. Specifically, if a $W/Z + j$ collision occurs at the same place as a second collision that yields soft leptons, the resulting final state is the same as our signal. The cross section for a DPS event is
\begin{equation}
\sigma_{tot} = \frac{\sigma_A\, \sigma_B}{\sigma_{\rm eff} }, 
\end{equation}
where $\sigma_A$ and $\sigma_B$ are the cross sections for the two separate processes, and $\sigma_{\rm eff}$ is an effective total cross section\footnote{This is only an approximation, as this expression neglects correlations among the two subprocesses and cannot hold in all corners of phase space.}. From Tevatron measurements~\cite{Abe:1997xk, Abazov:2009gc} and DPS theory (see Ref.~\cite{Berger:2009cm} and references therein), $\sigma_{\rm eff} \sim 12\, \mb$.  DPS has also been measured at the LHC~\cite{Aad:2013bjm,Gunnellini:2013nxa}, finding similar $\sigma_{\rm eff}$. The cross section for $W/Z + j$ is large, even after we apply $p_{T,j}$ and $\slashed E_T$ cuts, and there are many QCD and EW processes that generate soft leptons, so the rate for both processes occurring simultaneously is not obviously negligible. We must investigate further, looking at the various sources of soft leptons.
\begin{itemize}
\item One source of soft leptons is Drell-Yan (DY) production: $pp \to \ell^+\ell^- + X$. At $8\, \tev$, the leading order cross section after enforcing $p_{T, \ell} > 7\, \gev,\, |\eta_{\ell}| < 2.5$ is $\sim 1.1\, \nb$. The $K$-factor for DY is large, roughly $1.5$~\cite{Martin:2007bv, Hamberg:1990np}, but let us round that up to 2 to account for potential difference in acceptance between the LO and higher-order processes. Combining the DY rate with the cross section for $Z + j$: $\sigma(pp \,\to Z(\nu\bar{\nu}) + j), p_{T,j} > 100\, \gev, \slashed E_T > 100\, \gev \simeq 50\, \pb$ at 8 TeV, we find:
\begin{equation}
\sigma_{DPS} \simeq 50\, \pb \times \Big( \frac{2\, \nb}{12 \times 10^6\, \nb} \Big) \simeq 10^{-5}\, \pb = 10^{-2}\, \fb,
\end{equation}
a rate much smaller than either the signal or the dominant background.
\item QCD resonances are a second source of low-energy leptons, i.e $pp \to \Upsilon \to \ell^+\ell^-$. Because they are formed from colored constituents, these resonances have a large, strong-production sized cross section. For example, the LHC has measured $pp \to \Upsilon + X$ at 7 TeV to be $\sim 7.4\, \nb$~\cite{Khachatryan:2010zg}. However, the branching ratios for these resonances into leptons is rather small, $O(2\,\%)$. Doubling the 7 TeV rate and using $BR(\Upsilon \to \ell^+\ell^-) = 5\, \%$ to be conservative and account for the increase of energy from 7 TeV to 8 TeV, we find the net $pp \to \Upsilon + X \to \ell^+\ell^- + X \simeq 0.74\, \nb$, less than the DY number quoted above and without considering any kinematic cuts on the leptons. We saw that the DY rate was an order of magnitude too small to be a concern, so this smaller DPS source is certainly negligible. 
\item The final source of low-energy leptons we consider is continuum heavy quark production $pp \to \bar b b, \bar c c + X$ followed by semi-leptonic decays. The rate for these processes is orders of magnitude larger than resonance production, but the final states they produce are much dirtier. To account for the fact that the leptonic decay products only inherit a fraction of the parent quark's momenta and must lie within the tracker rapidity in order to be recognized, we estimate the continuum heavy quark production by $pp \to \bar b b + X$ with cuts $p_{T,b} > 20\, \gev,\, |\eta_b| < 2.5$. Including a factor of two for charm quarks and another factor of two to (conservatively) account for higher order effects, we have $pp \to \bar bb, \bar cc + X \simeq  4\times 10^3\,\nb$ (8 TeV). In order to fake our signal, these heavy quarks must decay to isolated leptons. This can either be thought of as a one step process -- the $b/c$-jet faking a lepton, or as two steps -- semi-leptonic decay of a $b/c$ quark that happens to pass all lepton isolation criteria. Assuming a very conservative $b/c$ fake rate of $1\%$ (equivalent to an isolation fake of $\sim 10\%$ on top of the $\sim 10\%\, BR(b/c \to \ell + X)$), the net cross section drops to $0.4\, \nb$, again below the DY rate. Taking a more aggressive fake rate and incorporating all kinematic cuts, the rate will drop further.
\end{itemize}
We conclude that, unless $\sigma_{\rm eff}$ is significantly smaller than current measurements indicate, DPS is a negligible background for the final state configurations we are focused on. It is worth pointing out that the hard jet and $\slashed E_T$ requirements  play a key role in suppressing these potentially hazardous backgrounds. Reducing the jet (and $\slashed E_T$) cut to $50\, \gev$, the signal and dominant background would approximately triple, however the DPS background would increase by a factor of 6 since the cross section for $Z+j$ is a steeper function of the jet $p_T$ than diboson plus jet. More dramatically, had we been looking in trileptons, rather than two leptons plus a hard jet, the DPS background would be ($pp \to W^{\pm}+X)(pp \to $ soft leptons), approximately $200$ times larger. Finally, while the numbers we have quoted come from studies using 8 TeV LHC, we do not expect the conclusions to change for 14 TeV.\\

The final environmental background we consider comes from leptons that are lost due to detector inefficiencies or simply by falling outside the fiducial detector volume. By losing a lepton, a background that normally generates three or more leptons will fall into our final state. The most obvious candidates for this type of background are $W(\ell\nu)Z(\ell\ell) + j$ and $W(\ell\nu)\gamma^*(\ell\ell) + j$. To test the size of this contribution we generated $pp \to W(\ell\nu)Z/\gamma^*(\ell\ell)+ j$ and applied the analysis chain described earlier. The $m_{\ell\ell}$ distribution for these backgrounds is shown in Fig.~\ref{fig:minor_bg}, where we also plot $pp \to j + \ell\ell + \nu\bar{\nu}$ for comparison. The lost-lepton background is clearly negligible. 

\subsection{Other lepton multiplicities}
\label{sec:other}

So far we have concentrated on the two lepton signal, though other lepton multiplicities may be useful in different regimes of the electroweakino parameter space \cite{Giudice:2010wb}. A natural question to ask at this point is: what are the prospects for $pp \to j + \slashed E_T$ plus $n \not= 2$ leptons? If we require only one lepton, the final state is $pp \to j + \met + \ell $ and we face an immense background from $W(\ell \nu) + j$. At $8$ TeV, $\sigma(W^{\pm}(\ell\nu) + j), p_{T,j} > 100\, \gev, |\eta_j| < 2.5$ and $\slashed E_T > 100\, \gev$, the cross section is $\sim 40\,\pb$,  over five hundred times larger than the signal. In the $W+j$ events, the transverse mass formed from the lepton and the neutrino $m^2_T = 2\,\met p_{T,\ell}(1-\cos(\phi_{\met-\ell}) )$, has a kinematic edge at  $M_W$, so by requiring $m_T \gg M_W$ one can dramatically reduce this background. However, the signal we are after also typically has small $m_T$, so a large $m_T$ cut is ineffective. Since the $\met$ may be large in the signal and there is no kinematic constraint on the $\ell-\slashed E_T$ system (such as a parent particle), one might expect the signal to have large $m_T$. The reason for small signal $m_T$ is that the signal leptons -- who's $p_T$ enter directly into the transverse mass definition --  are soft. Given the size of the $W + j$ background and the lack of handles to distinguish it from the signal, this possibility looks very challenging.

The three-lepton signal, $p p \rightarrow j + \slashed E_T$ plus $n \ge 3$ leptons,  does not suffer from large backgrounds, however after paying the price of a hard ISR jet and leptonic branching ratios, the rate is quite low. In addition, each extra lepton in the event must pass a minimum $p_T$ cut, so the more leptons present the more times we pay an efficiency price. Looking for signals with two same-sign leptons (SSL) is another interesting, low-background option, though for the spectra we are interested in the only source of SSL is trilepton events. Since the SSL signal $pp \to j + \slashed E_T + \ell^{\pm}\ell^{\pm}$ only requires two leptons rather than three, the lepton efficiency is higher than for the trilepton final state. Therefore, the rate may be high enough to be viable in some regions of parameter space the SSL, though a thorough study of this possibility is necessary.

\section{Hunting Degenerate Higgsinos:  Monojet Limits}
\label{sec:monoj}

We now turn to hunting for quasi-degenerate Higgsinos with the most difficult case, where $M_1, M_2$ are so large that the splitting among the Higgsinos is $\lesssim $ few GeV and the particles emitted in the decay cascades are so soft that they cannot pass ATLAS/CMS object identification requirements, let alone the trigger requirements. For all intents and purposes, the entire Higgsino sector is invisible in this case; $pp \to \chi^{+}_1\chi^{-}_1,\, pp \to \chi^{+}_1\chi^{0}_1, pp \to \chi^{+}_1\chi^{0}_2$ all look the same and can be combined into a single process $pp \to \chi\chi$. The degree of degeneracy is limited by the fact that the Higgsino decays need to be prompt in order for $pp \to \chi\chi+j$ to mimic a monojet signal. If the Higgsino decay lengths become macroscopic, additional search strategies that rely on displaced tracks or stubs can be used~\cite{Aad:2013yna}. The requirement of prompt decays sets a lower limit on the inter-Higgsino splitting of roughly $\sim 0.3\, \gev$.

\begin{figure}[t!]
\centering
\includegraphics[width=0.45\textwidth]{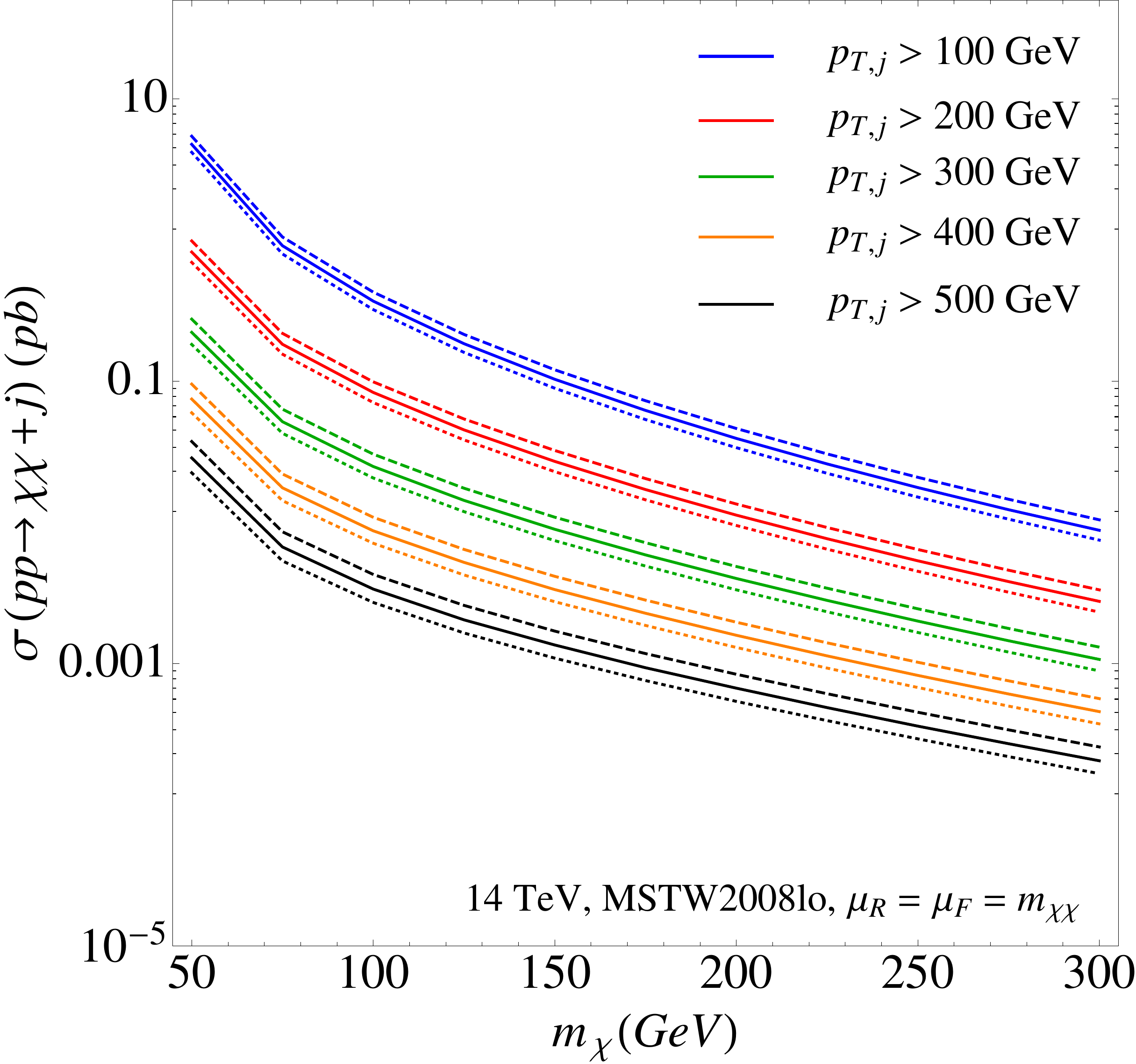}
\includegraphics[width=0.45\textwidth]{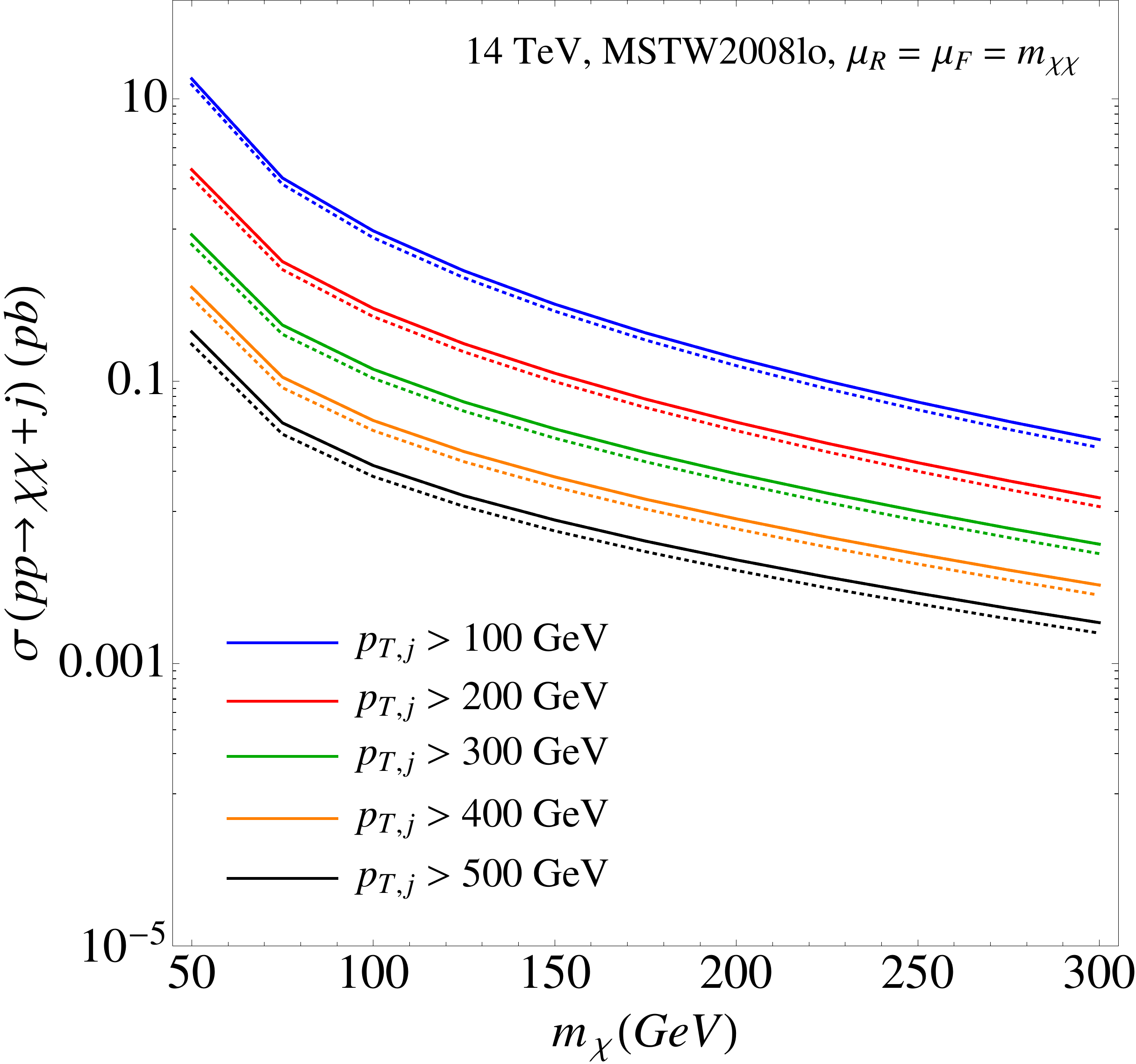}
\caption{Leading order Higgsino plus jet cross sections at 8 TeV (left) and 14 TeV (right) as a function of the Higgsino mass for several different jet $p_T$ cuts. The factorization/renormalization scale was taken to be $m_{\chi\chi}/\,0.5\,m_{\chi\chi}/\,2\,m_{\chi\chi}$ in the solid/dashed/dotted lines of each color, and all calculations use MSTW2008lo parton distribution functions.  Note the ``kink'' in the plot near $m_{\chi} = 70$~GeV is due merely to numerical sampling.}
\label{fig:higgsinoxsec}
\end{figure}

Pair production of degenerate Higgsinos looks just like DM pair production, except the mediator $\gamma, W, Z$ is light, the couplings are EW strength, and the Lorentz structure is dictated by the weak interaction: vector-vector and axial-axial  operators only. To get an idea for the typical Higgsino plus jet rates, the cross sections for $\sigma(pp \to \chi\chi + j)$ for several different jet $p_T$ are shown in Fig.~\ref{fig:higgsinoxsec} as a function of the Higgsino mass.
The signal cross sections were calculated assuming an exactly degenerate multiplet of Higgsinos, i.e. the four states pair up into two Dirac fermions, one charged and one neutral. These cross sections were calculated at leading order using MSTW2008lo~\cite{Martin:2009iq} parton distribution functions and a factorization/renormalization scale of $\mu^2 =  m^2_{\chi\chi}$, the invariant mass of the Higgsino pair system.

We cannot rescale bounds from existing monojet searches since Higgsino pair production proceeds through a light mediator ($W/Z$) rather than a contact interaction. In order to reinterpret CMS/ATLAS monojet bounds in terms of Higgsino pair production, we redo the existing analysis on Higgsino plus jet events and compare with the observed number of events in each bin.  However, because Higgsinos are not produced from a contact interaction, we do not need to worry about large jet-$p_T$ cuts invalidating our effective theory.

While the details of the CMS and ATLAS monojet searches differ slightly, they have the same basic strategy: clean events containing a single hard jet and substantial missing energy, divided into various $p_{T,j}$ and $\slashed E_T$ bins, then compared with the standard model expectation in each bin. The standard model background comes predominantly from $W(\ell \nu) + j$ and $Z(\nu \bar{\nu}) + j$ events. To get a rough idea of the current bounds, we apply the same analysis cuts (shown in full detail in Appendix~\ref{app:search}) the experiments use to parton-level Higgsino plus jet events. Assuming the states are sufficiently degenerate, the Higgsino plus jet cross section and the cut efficiency only depend on the Higgsino mass $m_{\chi}$. The cross section and cut efficiency $\epsilon$, combined with the luminosity, gives us the $s_i$, the total number of Higgsino events in a particular signal region. 
\begin{equation}
s_i(m_{\chi}) = \mathcal L \times \sigma( pp \to \chi\chi+j)(m_{\chi}) \times \epsilon(m_{\chi})
\end{equation}
By comparing $s_i(m_{\chi})$ to $s_{i,95}$, the 95\% CL limit on the number of allowed in a particular channel, we get a bound on $m_{\chi}$. The 95\% CL limit is calculated as
\begin{equation}
0.05 = \frac{ \int \delta b_i \text{Gaus}(\delta b_i)\times \text{Pois}(n_i | b_i (1 + \delta b_i) + s_{i,95} ) }{\int \delta b_i \text{Gaus}(\delta b_i)\times \text{Pois}(n_i | b_i (1 + \delta b_i) )},
\end{equation}
where $n_i$ is the number of observed events in bin $i$, $b_i$ is the expected SM contribution to that bin. The number of background events is modulated by a Gaussian with width $\delta b_i$, where $\delta b_i$ is the fractional uncertainty in bin $i$ quoted by the experiment. This modulation is done in an effort to incorporate the effects of systematic uncertainties, which are often large in monojet searches. To derive the expected limit on $s_i$, rather than the observed limit, one would replace $n_i$ with $b_i$ in the formula above.

Reinterpreting the limits from ATLAS/CMS latest monojet searches~\cite{atlasmono, cmsmono} we quote the most stringent {\em expected} limit on $m_{\chi}$:
\begin{align}
\text{ATLAS}:\,  &\, m_{\chi} > 73\, \gev\, (\text{SR3}) \nonumber \\
\text{CMS}:\,  &\, m_{\chi} > 80\, \gev\, (\text{SR5} )
\end{align}
The observed limits are slightly different:
\begin{align}
\text{ATLAS}:\,  &\, m_{\chi} > 103\, \gev\, (\text{SR4}) \nonumber \\
\text{CMS}:\,  &\, m_{\chi} > 73\, \gev\, (\text{SR5} ),
\end{align}
presumably originating from a slight downward (for ATLAS) or even slighter upward  (for CMS) fluctuation in the data. Full details of the bounds in all signal regions are shown in Appendix~\ref{app:search}. This study was done by taking the lowest-order cross section and using parton-level events. Incorporating parton showering, hadronization and detector effects will only loosen the bound, so we believe our estimate is optimistic. It would be interesting to see more detailed monojet studies including these details, but this is best left to the ATLAS and CMS collaborations. 

These limits are no better than the LEP bound of $\sim 103\,\gev$~\cite{lepchargino_lim}. While it is likely that the experimental collaborations will exceed our estimates once further studies help reduce the systematic uncertainties on backgrounds, additional search channels will only improve the situation. 

\section{Discussion}
\label{sec:disc}

In this paper we presented a novel search for electroweakinos $\chi$ based on $\chi\chi$ production in association with hard initial-state radiation. This search targets regions of supersymmetry parameter space where $\mu \ll M_1, M_2$, implying that the lightest electroweakino states are predominantly Higgsino and are quasi-degenerate (splitting of $O(m^2_Z/M_1, m^2_Z/M_2) \sim 5-50\, \gev$). This electroweak spectrum is characteristic of natural supersymmetry, where the role of the Higgsino mass parameter on the Higgs sector and EWSB demands that $\mu$ is at or near the weak scale.

The production of Higgsinos in association with a hard jet has a smaller rate than producing Higgsinos alone, but it comes with extra handles. In particular, the ISR jet can be used for triggering, which allows us to significantly relax the energy requirements on other final-state particles, making the $\chi\chi+j$ search sensitive to small inter-Higgsino splittings $\delta m_{\chi} < m_W$.  We focused on the final state with two leptons, $pp \to j + \slashed E_T + \ell\ell$. This final state has a priori large backgrounds, but we find these can be reduced with cuts. In particular, the $\bar t t$ background can be suppressed by requiring only one light-flavor jet, while $Z/\gamma^*(\tau+\tau^-) + j$ can be reduced by forming an effective $m_{\tau\tau}$ out of the $\ell\ell+\slashed E_T$ system and cutting out the $Z$-peak. The dominant remaining background, $W^+(\ell\nu)W^-(\ell\nu) + j$, typically generates harder leptons than the signal since the $W$ are on-shell, so we can find a signal-rich region by focusing on small $m_{\ell\ell}$. We have also performed a thorough study of the fake-lepton and double-parton backgrounds, finding them to be under control.  We find that evidence for light Higgsinos -- meaning right at the edge of the LEP II bound -- could be uncovered just by applying our search to the existing $8\, \tev$ data set. Specifically,  we can achieve $2\,\sigma$ evidence for $\mu = 110\, \gev$ if $M_1 = 1,\tev$ and $200\,\gev \lesssim M_2 \lesssim 750\,\gev$ or $M_2 = 1,\tev, M_1 \lesssim 550\,\gev$. Moving to 14 TeV, we find $5\,\sigma$ discovery is possible after $100\,\fbinv$ data for $\mu = 110\,\gev$ and nearly all $M_1, M_2 \lesssim 1\,\tev$. Increasing $\mu$ the signal rate decreases. However, we find $5\,\sigma$ discovery is possible (at $100\,\fbinv, 14\,\tev$) for $\mu \lesssim 140\, \gev$ with $2\,\sigma$ evidence possible for Higgsinos as heavy as $200\,\gev$ ($M_1 = M_2 = 500\, \gev$). Though it may take high energy and high luminosity, this search does shed light into a region traditional electroweakino searches simply cannot reach.

We emphasize that the search strategy we propose is complementary to the many other existing searches for gauginos.  For example, the trilepton plus $\slashed E_T$ search can be combined with our proposed search to cover a much larger portion of the full  electroweakino parameter. Additionally, the cuts and discrete choices in the $pp \to \chi\chi +j$ search (i.e. the number of leptons) should also be varied to extend the range of the search. For example, one could imagine decreasing the ISR jet requirement but increasing the $p_T$ requirement on the leading lepton. Diversifying even further, one can imagine an ensemble of electroweakino searches, spanning from monojet to trilepton with a variety of final state $p_T$-cuts and lepton multiplicities. 

We conclude with what we perceive to be the critical elements 
for the LHC collaborations to increase the signal and improve the search 
for quasi-degnerate Higgsinos:
\begin{itemize}
\item The lowest possible $j + \slashed E_T$ trigger to maximize the rate 
to search for off-line low $p_T$ leptons. 
\item A dedicated (or adapted) trigger for quasi-degenerate electroweakinos. 
For example $j + \ell + \slashed E_T$, with a low $p_T \lesssim 10$~GeV lepton, 
and a jet / missing energy threshold that is lower than the trigger without
the lepton. 
\item The lowest possible off-line lepton identification $p_T$ to maximize 
the number leptons that can be found in the triggered event samples. 
\item Fully explore the dilepton invariant mass into QCD resonance region, 
$m_{\ell\ell} < 12$~GeV, where the signal significance is maximized when
the Higgsinos are quasi-degnerate. 
\item Explore additional signals where Higgsinos are recoiling against 
other objects, i.e.\ $\gamma + \slashed E_T$, or  $W/Z + \slashed E_T$ 
signatures \cite{Gershtein:2008bf, Fox:2011pm}. \\
\end{itemize}

\emph{Note added:} As this paper was being completed, Ref.~\cite{Schwaller:2013baa} appeared that also  discussed strategies for electroweak production of compressed gauginos.  \\

\section*{Acknowledgments}
We thank J.P.~Chou, J.~Evans, S.~Gori, A.~Hook, and S.~Somalwar for useful discussions.
GDK thanks the Ambrose Monell Foundation for support while at the
Institute for Advanced Study.
ZH, GDK, and AM are supported in part by the US Department of Energy under 
contract number DE-FG02-96ER40969. ZH is also supported in part by DOE grant NO DE-FG02-13ER41986.  

\appendix
\section{Monojet Search details}
\label{app:search}

\subsection*{ATLAS monojet, $10\, \fbinv$, 8 TeV~\cite{atlasmono} }
\noindent
Trigger:
\begin{itemize}
\item $\slashed E_T > 80\, \gev$
\end{itemize}
Basic cuts:
\begin{itemize}
\item one jet with $p_{T,j} > 120\, \gev,\, |\eta_j| < 2.0$. If there is a second jet in the event with $p_T > 30\, \gev, |\eta| < 4.5$, veto the event
\item $\slashed E_T > 120\, \gev$
\item in events with a second (soft) jet, demand $\Delta \phi_{\slashed E_T, j_2} > 0.5$
\item veto any event with an isolated lepton satisfying $p_T > 20\, \gev, |\eta| < 2.47$ (electrons) or $p_{T} > 7\, \gev, |\eta| < 2.5$ (muons).
\end{itemize}
Analysis:
\begin{itemize}
\item Surviving events are binned into four signal regions: $\slashed E_T, p_{T,j} > 120\, \gev$, $> 220\, \gev$, $> 350\, \gev$, and $> 500\, \gev$.
\end{itemize}
\begin{table*}[t!]
\centering
\begin{tabular}{|c|c|c|c|} \hline
 Channel & Observed Limit & Expected Limit & Description \\  \hline
 ATLAS SR1 & $50\,\gev$ & $52\,\gev$ & $p_{T,j}, \slashed E_T > 120\, \gev$ \\
 ATLAS SR2 & $70\, \gev$ &  $69\, \gev$ & $p_{T,j}, \slashed E_T > 220\, \gev$\\
 ATLAS SR3 & $67\, \gev$ & $73\, \gev$ & $p_{T,j}, \slashed E_T > 350\, \gev$\\
 ATLAS SR4 & $103\, \gev$ & $63\, \gev$ & $p_{T,j}, \slashed E_T > 500\, \gev$\\ \hline
 \end{tabular}
\caption{Limits on the mass of a degenerate Higgsino multiplet coming from monojet searches from CMS and ATLAS}
\end{table*}

\subsection*{CMS monojet, $19.5\, \fbinv$, 8 TeV~\cite{cmsmono} }
\noindent
Trigger:
\begin{itemize}
\item 1 jet, $p_T > 120\, \gev$ or the combination of $\slashed E_T > 105\, \gev$ and 1 jet $p_T > 80\, \gev, |\eta| < 2.6$
\end{itemize}
Basic cuts:
\begin{itemize}
\item one jet with $p_{T,j} > 110\, \gev,\, |\eta_j| < 2.4$. If there is a second jet in the event with $p_T > 30\, \gev$, veto the event
\item $\slashed E_T > 250\, \gev$
\item in events with a second jet, demand $\Delta \phi_{j_1, j_2} > 2.5$
\item veto any event containing a lepton: $p_{T} > 10\, \gev, |\eta| < 2.5$ (electrons/muons) or $p_{T} > 20\, \gev, |\eta| < 2.3$ (taus)
\end{itemize}
Analysis:
\begin{itemize}
\item Surviving events are binned according to their $\slashed E_T$: $\slashed E_T > 250\, \gev,\, > 300\, \gev,\, > 350\, \gev,\, > 400\, \gev$, $> 450\, \gev,\, > 500\, \gev$ and $> 550\, \gev$.
\item Notice that only the $\slashed E_T$ is incremented in the binning, while the $p_{T,j}$ cut is constant.
\end{itemize}

 \begin{table*}[t!]
\centering
\begin{tabular}{|c|c|c|c|} \hline
Channel & Observed Limit & Expected Limit & Description \\  \hline
 CMS SR1 & $64\, \gev$ & $60\, \gev$ & $\slashed E_T > 250\, \gev$\\
 CMS SR2 & $68\, \gev$ & $63\, \gev$ & $\slashed E_T > 300\, \gev$\\
 CMS SR3 & $70\, \gev$ & $68\, \gev$ & $\slashed E_T > 350\, \gev$\\
 CMS SR4 & $71\, \gev$ & $71\, \gev$ & $\slashed E_T > 400\, \gev$\\
 CMS SR5 & $73\,\gev$ & $80\, \gev$ & $\slashed E_T > 450\, \gev$\\
 CMS SR6 & $71\, \gev$ & $74\, \gev$ & $\slashed E_T > 500\, \gev$\\
 CMS SR7 & $68\, \gev$ & $68\, \gev$ & $\slashed E_T > 550\, \gev$\\ \hline
\end{tabular}
\caption{Limits on the mass of a degenerate Higgsino multiplet coming from monojet searches from CMS and ATLAS}
\end{table*}


\end{document}